\documentclass[twocolumn,twocolappendix]{aastex701}

\usepackage{amsmath}
\usepackage{placeins}
\usepackage{float}
\usepackage{graphicx}
\usepackage{booktabs}

\begin{document}

\title{Predicted Spins of Galactic Satellite Systems and How to Observe Them}

\author[orcid=0009-0002-4568-2423]{Luke Y. Xia} 
\affiliation{Department of Physics and Astronomy, University of California Irvine, Irvine, CA 92697, USA}
\affiliation{Kavli Institute for Particle Astrophysics and Cosmology and Department of Physics, Stanford University, Stanford, CA 94305, USA }
\email[show]{lyxia@stanford.edu}    

\author[orcid=0000-0003-4298-5082]{James S. Bullock}
\affiliation{Department of Physics and Astronomy, University of Southern
California, Los Angeles, CA 90089, USA}
\email{jamesbul@usc.edu}

\submitjournal{ApJ}

\begin{abstract} We use dark matter only (Symphony)  and hydrodynamic (TNG50, FIRE-2, and Auriga) simulations to study the spins of satellite systems around Milky Way-like galaxies. Treating each subhalo as a point-like tracer, we find that the satellite system spin, $\lambda_{\rm sat}$, correlates with underlying dark matter halo spin, $\lambda_{\rm dm}$, with median ratio $\lambda_{\rm sat}/\lambda_{\rm dm} \simeq 1.64$.  The sample-wide average is $\langle \lambda_{\rm sat}\rangle =0.06$ and the halo-to-halo distribution is roughly log-normal with scatter $\sigma_{\ln \lambda} = 0.55$. We use sampling theory to derive closed-form expressions for the relationship between tracer counts and spin measurement bias; accounting for this, we find no dependence of $\lambda_{\rm sat}$ on subhalo mass. We show analytically and with mock surveys that the satellite spin projected onto the disk rotation axis, $\lambda_\parallel$, can be measured accurately by ``stacking" satellite systems with projected velocities around disks with observed rotation directions. Using the hydrodynamic runs, we predict $\langle \lambda_\parallel \rangle = 0.03$ and a related observable: the fraction of satellites with projected orbits that are prograde with respect to the disk, $f_{\rm pro}=0.53$. The uncertainty on either stacked mean depends inversely on the total number of satellites observed, $\propto 1/\sqrt{HN}$, where $H$ is the number of hosts and $N$ is the number of satellites per host.  For example, an idealized survey of 2000 host disks with one satellite velocity measurement each would allow us to measure both $\langle \lambda_\parallel \rangle$ and $f_{\rm pro}$ to within $\sim 0.01$. 
\end{abstract}


\section{Introduction}
\label{sec:introduction}

Among the longest-standing predictions of dark matter based cosmologies is that the specific angular momentum, $j_{\rm dm}$, of dark matter halos should be small, but non-zero, compared to the values required for rotational support \citep{Peebles1969,Doroshkevich1970,Peebles1971,White1984}. Although these predictions and the work that followed provided an important baseline for understanding angular momentum in galaxies \citep{Fall1980,Mo1998}, the spins of dark matter halos themselves are not currently  observable. In this paper, we present predictions for a related quantity: the host-centric specific angular momentum of satellite systems in halos, $j_{\rm sat}$, treating each satellite or subhalo as a point tracer.
We formulate the prediction in terms of a satellite halo spin parameter, $\lambda_{\rm sat}$, and demonstrate that its disk-aligned component, $\lambda_\parallel$, should be directly observable using stacking experiments with current and planned surveys. 

\citet{Peebles1971} was the first to introduce a dimensionless spin parameter for characterizing the degree of rotational support in halos: $\lambda_{\rm P} = j_{\rm dm} |E|^{1/2} G^{-1} M^{-3/2}$, where $j_{\rm dm}$ is the specific angular momentum of dark matter within the halo volume. Peebles, along with the other authors cited in the paragraph above, showed that tidal torque theory (TTT) in an expanding universe predicts that dark matter halos should have spins consistent with those needed to explain the sizes of disk galaxies\footnote{\citet{Hoyle1949} was the first to suggest nonspherical external gravitational fields torqued protogalaxies and gave rise to their angular momentum. See \citet{Efstathiou03} for a discussion of this work.}. Subsequently, \citet{Barnes1987} and \cite{Warren1992} used CDM N-body simulations to show that the halo-to-halo spin distribution is roughly log-normal, with median values only slightly below the analytic TTT predictions.  Later, \citet{bullock2001}  introduced a modified spin parameter that is easier to measure in simulations: $\lambda_{\rm dm} \equiv j_{\rm dm}/\sqrt{2} R_{\rm vir} V_{\rm vir}$.  The factor of $\sqrt{2}$ ensures that $\lambda_{\rm dm}   \simeq \lambda_{\rm P}$ for typical~\footnote{The equivalence is exact in the limit that dark matter halos are singular isothermal spheres.} halos at $z=0$.  

In the years that followed, larger simulations provided more clarity on halo spin predictions.  For example, \cite{Maccio2008} showed that while halo concentrations scale strongly with parameters such as $\sigma_8$ and $\Omega_m$, the dimensionless spin parameter distribution $P(\lambda_{\rm dm})$ is virtually universal across cosmologies. More recently, \cite{RodriguezPuebla2016} used Bolshoi--Planck and MultiDark--Planck simulations to show that while the energy-dependent spin parameter, $\lambda_{\rm P}$, decreases with redshift, the modified parameter, $\lambda_{\rm dm}$ remains nearly constant across cosmic time ($z = 0\text{--}8$). The consensus modern result is that the distribution of halo spins has a median $\lambda_{\rm dm} \simeq 0.035$, and, when fit to a log-normal distribution, has a scatter $\sigma_{\rm ln \lambda} \simeq 0.6$ \citep[or $\sigma_{\rm log_{10}} \simeq 0.26$;][]{Maccio2008,RodriguezPuebla2016,Zjupa2017}.

The bound structures that are accreted as subhalos or satellites of main halos are subject to the same tidal forces that spin up their hosts. 
\citet{Libeskind2012CosmicWeb} showed that
the orbital angular momenta of subhalos tend to align with the intermediate eigenvector of the cosmic-web velocity-shear tensor.  They further found that
the host-halo spin direction is correlated with the large-scale web geometry. In fact, accretion-based random-walk models provide a description of halo-spin acquisition that is complementary to TTT \citep{Vitvitska2002}. 
In particular, \citet{Benson2020} showed that
the distribution of halo spins can be reproduced when the angular momenta of infalling halos are allowed to correlate weakly with the spin of the growing host.  

Given this as background, we expect that present-day satellite populations,  when treated as point particles,  should have similar characteristic spin values as the dark matter particles of their hosts: $\lambda_{\rm sat} \sim \lambda_{\rm dm}$. However, we do not expect the match to be perfect.  Models of merging-halo orbital parameters emphasize the role of the pairwise velocity field and capture geometry \citep{Benson2021}. Further, subhalos evolve dynamically within host halos differently than do background dark matter particles \citep{ZentnerBullock2003,Green2021_SatGen2}. For example, when selected at fixed mass (as a proxy for a luminosity-limited sample), subhalo radial distributions are generally more extended \citep{nadler2023} than the background halo particles.  This suggests that satellite specific angular momenta ($\propto R_{\rm sat} V_{\rm sat}$) will tend to be slightly larger than that of the dark matter ($\propto R_{\rm dm} V_{\rm dm}$).

Past work has explored the angular momentum associated with substructure and found that it is substantial. \citet{Shaw2006} showed that subhalos in clusters have more orbital angular momentum per unit mass than the host as a whole and that their net angular momentum is preferentially aligned with the host spin. \citet{WarnickKnebe2006} found that approximately $60\%$ of subhalos are intrinsically prograde relative to the host halo, with the apparent fraction reduced to approximately $53\%$ in projection. Hydrodynamical simulations likewise show a preference for prograde satellite orbits and alignment between satellite systems and the outer dark-matter halo, although the satellite--galaxy alignment can be weakened by misalignment between the central disk and the outer halo \citep{Deason2011}.

More recently, \citet{Mezini2023} demonstrated that excluding subhalos reduces the median $\lambda_{\rm dm}$ by approximately $10\%$ for Milky Way--mass halos and $30\%$ for cluster-mass halos, showing that substructure makes a disproportionate contribution to the angular momentum budget. \citet{Mezini2025} extended this work to show that the net orbital angular momentum of the subhalo population is well aligned with that of the smooth host component. Importantly, this alignment persists when the subhalo angular momentum is constructed from the positions and velocities of the subhalo centers (as point particles), a definition closely related to the point-satellite statistic considered here.

While most previous studies related to the angular momentum of subhalos have focused on the mass-weighted angular momentum contribution of substructure, our work instead characterizes the distribution of an equal-weight, point-tracer spin parameter, $\lambda_{\rm sat}$, with an eye towards observability through stacking satellite systems that have central disks. The existence of a disk provides the ability to orient the stack with respect to the disk spin direction. As discussed below, such an experiment allows us to directly measure the satellite spin projected along the direction of the disk angular momentum, $\lambda_\parallel = \lambda_{\rm sat} \cos(\theta_{\rm disk, sat})$. Our simulation analysis focuses on Milky Way--mass hosts because this is a mass range with both a high fraction of rotating disks and a fair population of visible satellites. In particular, the hydrodynamic simulations that we use are restricted to have central disks; this allows us to predict $\lambda_\parallel$ and thus connect to a direct observable.

The connection between central disk galaxy and halo angular momentum is not one-to-one \citep{Deason2011,SGK2018}, but hydrodynamical simulations often find a statistically significant relationship between them.  In the IllustrisTNG simulations, for example, 
 late-type galaxies
typically retain $50$--$60\%$ of the specific angular momentum of their
host halos \citep{RodriguezGomez2022AngularMomentum}.  These results motivate the use of disk rotation as an imperfect observational proxy for the angular-momentum direction of the host satellite system; we demonstrate below that the correspondence is far from perfect, but also that the alignment is a predicted quantity. 

\citet{Zaritsky1997} were among the first to perform a stacking experiment to study satellite dynamics with respect to disk rotation. Using 69 isolated spiral galaxies with 115 satellites, they found a preference for satellites to rotate in the same direction as the primary disks. Although they did not measure a stacked spin parameter, a related quantity, $f_{\rm pro}$, the fraction of satellites that orbit in a prograde direction with respect to the host disk can be inferred from their Figure 8, which shows $f_{\rm pro} \simeq 0.54$. A similar analysis using 43 primaries with 76 satellites by \citet{Azzaro2006} reported  $f_{\rm pro} = 0.61$.  Later, \citet{HerbertFort2008} used a total of 169 satellites around 118 disks to measure $f_{\rm pro}=0.53$.  Relying on a larger sample of 635 satellites around 215 hosts, \citet{HwangPark2010} found no clear signal for co-rotation, and in fact reported a mean projected prograde fraction consistent with isotropy $f_{\rm pro} = 0.50$. In contrast, the more recent work of \citet{Wang2024} looked at an even larger sample of 580 central disks, each with one or two satellites, and concluded that satellite orbits were preferentially prograde with respect to their host disks at $\sim 3 \sigma$, but did not report a projected prograde fraction. Data of this kind is precisely what is needed to test the predictions we explore below.

Our analysis proceeds as follows. In the next section we provide an overview of the simulations used in this analysis and define the satellite spin parameter explicitly. Section \ref{sec:finite-n-theory} is devoted to finite-sampling theory for point-tracer spin measurements, and presents analytic expressions for how tracer counts map to RMS errors in spin measurements and directional misalignment measurements.  In Section \ref{sec:satspins} we provide predictions for point-like satellite 3D spin distribution, $\lambda_{\rm sat}$, explore the relation to dark matter halo spin, and study satellite spin directional alignment with host disks. With these theoretical quantities in hand, in Section \ref{sec:observational} we introduce an observable that can be used to directly measure the average satellite spin projected in the direction of disk rotation, $\lambda_\parallel$, provide predictions for both $\lambda_{\parallel}$ and $f_{\rm pro}$,  and derive closed form expressions for the expected precision on the measurements of these quantities as function of host and satellite count in an idealized survey. Section \ref{sec:mock-surveys} presents results of mock surveys that validate the derivations in Section \ref{sec:observational}. We conclude and provide discussion in Section \ref{sec:Conclusions}.

\section{Methods}
\label{sec:methods}

\subsection{Simulations and Host Selection}
\label{subsec:simulations}

We use four sets of publicly-available simulations: Symphony \citep{nadler2023}, TNG50 \citep{Nelson_2019,Pillepich2019TNG50, pillepich2024}, FIRE-2 m12/ELVIS zooms \citep{GarrisonKimmel2019,Wetzel2023,Wetzel_2025_DR2}, and Auriga \citep{Grand2017,Grand2024}. 
As detailed below, this group includes one dark-matter-only simulation, and 3 hydrodynamic suites with full galaxy formation modeled with different prescriptions and resolutions.  Table \ref{tab:simulation-suites} summarizes our sample and provides the dark matter particle mass resolution of each simulation. Our final sample includes a total of 176 hosts, 131 of which are hydrodynamic.

We restrict the hydrodynamic sample to consist of halos hosting central disk galaxies. We identify the disk and define the associated rotation axis using stars that are younger than 500 Myr within $0.1R_{\rm vir}$. 
This choice allows enough stars for a robust estimate of the star-forming disk’s rotation axis, and also traces populations that are somewhat intermediate between what might be traced by $H\alpha$ (ionized gas powered by younger stars), CO (molecular gas tracing similar age stars~\footnote{\citet{Quirk2019} found that bright-AGB populations in M31 ($ \sim 400\,\mathrm{Myr}$), have almost identical kinematics to the CO-emitting molecular gas.}), or inner HI (similar to these stars in our simulations).  
In order for the galaxy to qualify as a disk we use the diagnostic from \citet{Sales2012} and demand that the fraction of kinetic energy in rotation must obey $k_{\rm rot}>0.5$, where $k_{\rm rot} = \langle v_\phi^2\rangle_{m}/\langle v^2\rangle_{m}$, with $m$ denoting mass weighting among these young stars and $\phi$ denoting the angular direction of disk spin. 

\begin{table*}[t!]
\centering
\caption{Simulation host samples used in this work. The virial-mass column gives the median followed by the full range in parentheses. Other host quantities are medians unless indicated. Here $m_{\rm DM}$ is the dark-matter particle mass, $\sigma_{\ln\lambda_{\rm dm}}$ is the (natural) logarithmic spin scatter, and $Q_1$ and $Q_3$ denote the 25th and 75th percentiles of $\kappa_{\rm dm}$.}
\label{tab:simulation-suites}
\small
\renewcommand{\arraystretch}{1.06}
\setlength{\tabcolsep}{2pt}
\begin{tabular*}{\textwidth}{@{\extracolsep{\fill}}llrrrrrrrr@{}}
\toprule
Suite & Physics & $N_{\rm host}$ &
\shortstack{med (range) $M_{\rm vir}$\\($10^{12}\,M_\odot$)} &
\shortstack{med $R_{\rm vir}$\\(kpc)} &
\shortstack{med $V_{\rm vir}$\\(km s$^{-1}$)} &
\shortstack{$m_{\rm DM}$\\($10^5\,M_\odot$)} &
\shortstack{med\\$\lambda_{\rm dm}$} &
$\sigma_{\ln\lambda_{\rm dm}}$ &
\shortstack{$Q_1$/med/$Q_3$\\$(\kappa_{\rm dm})$} \\
\midrule
Symphony & DMO zoom-ins & 45 & 1.22 (1.13--1.40) & 278 & 137 & 4.03 & 0.030 & 0.70 & 37.9/81.5/173.4 \\
TNG50 & Hydro box & 93 & 1.13 (0.75--1.93) & 275 & 133 & 4.54 & 0.034 & 0.58 & 31.9/58.1/134.5 \\
FIRE-2 m12/ELVIS & Hydro zoom-ins & 12 & 1.25 (0.98--1.88) & 283 & 138 & 0.274 & 0.039 & 0.54 & 31.5/49.9/69.5 \\
Auriga & MHD zoom-ins & 26 & 1.45 (1.11--2.24) & 298 & 144 & 3.16 & 0.038 & 0.63 & 32.8/49.1/134.8 \\
\midrule
All suites & Mixed & 176 & 1.23 (0.75--2.24) & 280 & 137 & varied & 0.033 & 0.62 & 32.3/64.2/152.1 \\
\bottomrule
\end{tabular*}
\end{table*}

For all simulations analyzed, we adopt the Bryan--Norman convention \citep{bryan1998} for $M_{\rm vir}$ and $R_{\rm vir}$. We focus on Milky Way size host halos with $7.5 \times10^{11}<M_{\rm vir}/M_\odot<2.25\times10^{12}$. With this and the disk-selection criteria, our sample of host halos in each suite is summarized as follows.

 \textbf{Symphony:} We use all 45 Milky Way--mass dark-matter-only zoom-ins \citep{Mao2015}, which have a narrow range of virial masses: $M_{\rm vir} = 1.1-1.4 \times 10^{12}$ M$_\odot$. Host and subhalo phase-space coordinates come from the ROCKSTAR-based Symlib catalogs \citep{Behroozi2013Rockstar} and we keep subhalos with $r <R_{\rm vir}$. 

 \textbf{TNG50:} From the 198 MW/M31 analogues in the public TNG50 catalog \citep{pillepich2024}, we take the 137 systems satisfying \texttt{FlagCentral}$=1$ and \texttt{FlagM200c}$=1$. Although the initial selection uses an $M_{200c}$ flag, we adopt the catalog's Bryan--Norman virial quantities, \texttt{Group\_M\_TopHat200} and \texttt{Group\_R\_TopHat200}, for $M_{\rm vir}$ and $R_{\rm vir}$ throughout our analysis. Satellites require \texttt{SubhaloFlag}$=1$, \texttt{SubhaloParent}$=0$, and $r<R_{\rm vir}$, so nested sub-subhalos are not counted independently. After imposing our disk criteria and mass criteria, we are left with 93 hosts with virial mass range $M_{\rm vir} = 0.75-1.9 \times 10^{12}$ M$_{\odot}$.

 \textbf{FIRE-2 m12/ELVIS zoom-ins:} Our highest-resolution hydrodynamic sample contains twelve hosts with virial masses $M_{\rm vir} = 1-1.9 \times 10^{12}$ M$_{\odot}$ that pass the mass and disk criteria, including both isolated and Local-Group-like pairs.  We select ROCKSTAR subhalos within $R_{\rm vir}$ that belong directly to the main host, excluding satellites of other satellites. 
 
 \textbf{Auriga:} We use 26 level-4 Original Auriga MHD zoom-ins with $M_{\rm vir} = 1.1 - 2.2 \times 10^{12}$ M$_{\odot}$ that pass the host-mass and disk cuts. Satellites are direct SUBFIND children of the central galaxy with $r<R_{\rm vir}$.

\subsection{Satellite / Subhalo Selection}
\label{subsec:satellite-selection}

The required particle limit for modern halo finders to identify the position and bulk motions of a subhalo is fairly minimal, $N_{\rm res} \simeq 20$  \citep{Knebe2011,Onions_2012,Knebe2013,Avila14}.  
Since we treat subhalos as point particles, we do not need accurate masses or other internal properties, which would require much larger particle counts \citep[e.g.,][]{nadler2023}. We do not even need a complete sample at fixed mass, assuming that the sample is unbiased. Instead, our analysis benefits from having as many tracer subhalos as possible for sampling the angular momentum distribution. 

For our fiducial sample, we include all subhalos in the Symphony, TNG50, FIRE-2, and Auriga catalogs with recorded dark matter masses $M_{\rm dm}>10^7$ M$_\odot$.  This corresponds to particle counts of 25, 22, 365, and 32,  respectively. Note that while positions and bulk motions of overdensities identified with $\sim 20$ particles should be accurate,  their recorded masses are almost certainly under-estimated. That is, the choice of $10^7$ M$_\odot$ as a mass cut for Symphony, TNG50, and Auriga should really be thought of as particle-count cuts and {\em not} a claim that the catalogs are complete to a current subhalo mass of $10^7$ M$_\odot$.  Only the FIRE-2 simulations are actually complete to this mass.  What we {\em do} assume is that the identified subhalos in each catalog provide a fair subsample for characterizing the underlying satellite angular momentum properties.  This assumption is justified later in the paper, where we use higher mass cuts  to show that the results are consistent with being independent of subhalo mass once finite-sampling bias is included. We also find that the median (mean) satellite spin measurements from the FIRE-2 halos, which {\em are} complete, are within $2\%$ ($9\%$) of the full sample value, and are intermediate compared to the results from the other three simulation samples. For context, the median (mean) {\em dark-matter halo} spin of the FIRE-2 halos we use differs by $21\%$ ($2 \%$) of the full sample value.

\section{Finite-sampling and spin measures}
\label{sec:finite-n-theory}

\subsection{Magnitude Bias}
\label{subsec:finite-n-magnitude}

As discussed by \citet{Bett2007} and \citet{Benson2017}, 
random sampling adds an effective noise vector to the measurement of specific angular momentum vectors, and this yields a vector length that always increases even though the direction of the noise is random.
In order to see this, and to understand the magnitude of the effect, let $\boldsymbol j_i=\boldsymbol r_i\times\boldsymbol v_i$ be the specific angular momentum of tracer $i$.  The full population contains $N_{\rm tot} \gg 1$ tracers and has vector mean
\begin{equation}
\overline{\boldsymbol j}=\frac{1}{N_{\rm tot}}\sum_{i=1}^{N_{\rm tot}}\boldsymbol j_i,
\end{equation}
with magnitude $|\overline{\boldsymbol j}|$, and an associated spin parameter
\begin{equation} \lambda=\frac{|\overline{\boldsymbol j}|}
    {\sqrt{2}R_{\rm vir}V_{\rm vir}}.
    \label{eq:full-vector-mean}
\end{equation}
Now select $N$ of the $N_{\rm tot}$ tracers uniformly without replacement.  This produces a subsampled specific angular momentum mean $\overline{\boldsymbol j}_N$, with magnitude $|\overline{\boldsymbol j}_N|$, encoded in a subsampled spin $\lambda_N$.  We can write the new subsampled vector mean as
\begin{equation}
    \overline{\boldsymbol j}_N = \overline{\boldsymbol j} + \boldsymbol{\epsilon},
\end{equation}
where $\boldsymbol{\epsilon}$ represents the sampling error. The expected value of the vector sampling error is zero, $\mathbb{E}[\boldsymbol{\epsilon}]=0$, but the associated {\em magnitude} of the sampling error vector is not $\mathbb{E}[|\boldsymbol{\epsilon}|^2] \ne 0$. This means that the magnitude of the subsampled vector increases on average:
    $|\overline{\mathbf{j}}_N|^2  =|\overline{\mathbf{j}}|^2+2\overline{\mathbf{j}}\cdot\boldsymbol{\epsilon}+|\boldsymbol{\epsilon}|^2 \Longrightarrow |\overline{\mathbf{j}}|^2+|\boldsymbol{\epsilon}|^2$.
More formally, the expectation value of the magnitude squared is
\begin{equation}
    \mathbb{E}\!\left[|\overline{\mathbf{j}}_N|^2\right] = |\overline{\mathbf{j}}|^2+\mathbb{E}\!\left[\left|\overline{\mathbf{j}}_N-\overline{\mathbf{j}}\right|^2\right].
\end{equation}
The second term is the expected variance of a sample mean with finite-population correction, which is familiar from finite-sampling theory \citep[see, e.g.,][Sect.~7.3.1]{Rice2007}:
\begin{equation}
    \mathbb{E}\!\left[ |\overline{\boldsymbol j}_N-\overline{\boldsymbol j}|^2\right] = \frac{N_{\rm tot}-N}{N(N_{\rm tot}-1)} \left(\left\langle|\boldsymbol j|^2\right\rangle -|\overline{\boldsymbol j}|^2\right).
    \label{eq:sample-mean-variance}
\end{equation}
Note that the variance scales as
$1/N$ for $N \ll N_{\rm tot}$ and vanishes when $N = N_{\rm tot}$, as would be intuitively expected.
We can now define
\begin{equation}
    R_N\equiv\frac{\lambda_N}{\lambda} =\frac{|\overline{\boldsymbol j}_N|}{|\overline{\boldsymbol j}|}, \qquad \kappa\equiv \frac{\langle|\boldsymbol j|^2\rangle} {|\overline{\boldsymbol j}|^2}-1,
    \label{eq:kappa-definition}
\end{equation}
and write the expected squared ratio of subsampled spin to ``true" spin in closed form
\begin{equation}
    \boxed{\mathbb{E}[R_N^2] =1+ \frac{1}{q}; \hspace{.1in} q\equiv \frac{N(N_{\rm tot}-1)}{\kappa(N_{\rm tot}-N)}.}
    \label{eq:finite-n-exact}
\end{equation}

Here, $q \to N/\kappa$ for $N \ll N_{\rm tot}$ is a quantity that we refer to as the {\em tracer quotient}. When the tracer quotient is large, the sampling bias is small. It depends on both the number of tracers, $N$, and on the quantity $\kappa$, which we refer to as the angular-momentum {\em contrast}.  The contrast, $\kappa$, is a halo-specific quantity that compares the typical angular momentum of one tracer particle to the (much smaller) net angular momentum of the whole system.  The larger the $\kappa$, the larger $N$ needed for an accurate measurement.

\begin{figure}[t!]
\centering
\includegraphics[width=0.49\textwidth]{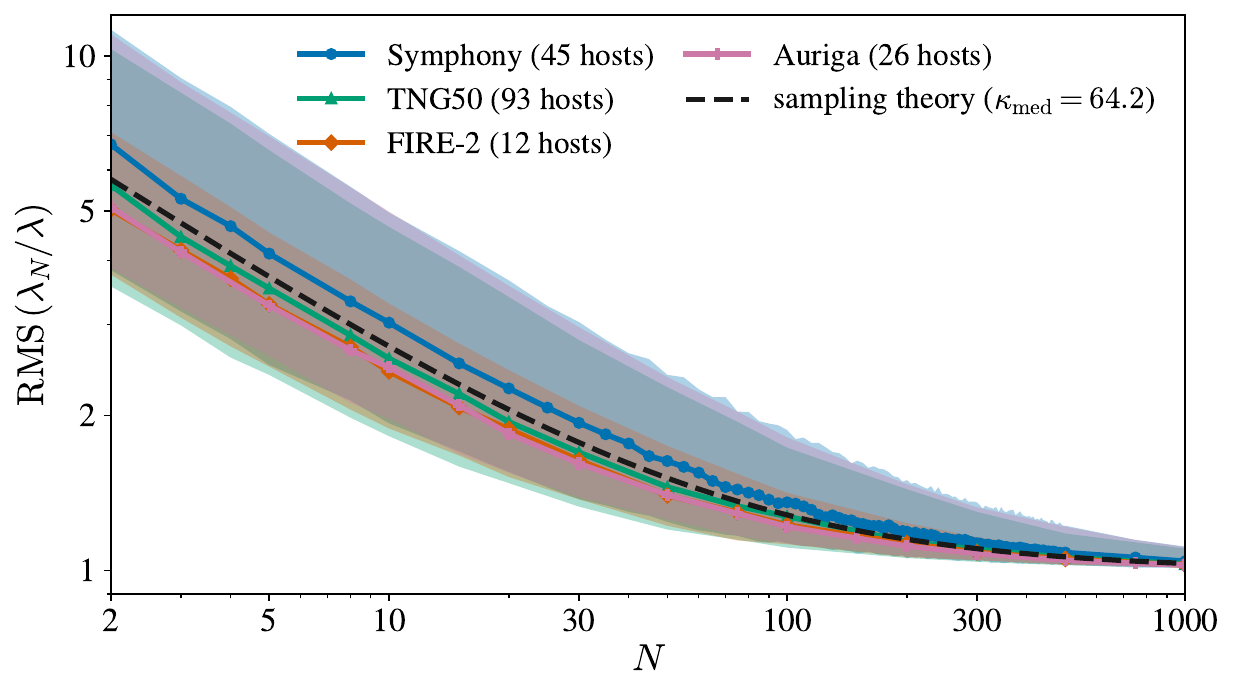}
\caption{How limited particle count, $N$, biases halo spin measurements, $\lambda_{\rm dm}$. Shown is the RMS ratio of the spin $\lambda_N$ measured using $N$ subsampled particles to the ``true" spin $\lambda$ (with $N_{\rm tot} \sim 10^6$) as a function of $N$.
For each host and $N$, we measure $\lambda_N$ using 2000 uniform draws without replacement.  Colored curves are suite medians and colored bands span the 16th--84th percentiles across hosts.  The black dashed curve is the analytic prediction given by Equation~\ref{eq:finite-n-rms} evaluated at the combined host median $\kappa=64.2$.}
\label{fig:finite-n-vs-n}
\end{figure}

What is a typical value for $\kappa$? A reasonable estimate for the typical angular momentum of an individual dark matter particle is $|\boldsymbol j|^2 \sim (0.5 R_{\rm vir} V_{\rm vir})^2$ \citep{vandenBosch1999}. This is large compared to the {\em net} angular momentum of all the particles, $|\overline{\boldsymbol j}|^2$, so we expect $\kappa$ to be significantly larger than unity:
\begin{equation}
    \kappa_{\rm dm} \sim \frac{(V_{\rm vir} R_{\rm vir})^2}{4 |\overline{\boldsymbol j}|^2} \sim \frac{1}{8 \lambda^2} \sim 50.
\end{equation}
Measured directly in our simulation sample of hosts, the median is $\kappa_{\rm dm}=64.2$, with a 16th--84th percentile interval of $25.3$--$217.5$.

It is useful to write the expected root-mean-square (RMS) enhancement factor due to finite-sampling as $F(q)$:
\begin{equation}
    F(q)\equiv
    \left\langle R_N^2 \right\rangle^{1/2{}}
    =\sqrt{1+\frac{1}{q}} \simeq \sqrt{1 + \frac{\kappa}{N}}.
    \label{eq:finite-n-rms}
\end{equation}
 From this we see that particle counts $N \lesssim \kappa$ will produce significant biases in spin estimates. Achieving less than a 10 percent bias would require $\kappa/N \lesssim 0.2$, or, equivalently, more than $N \sim 250$ particles for $\kappa = 50$. This is consistent~\footnote{\citet{Benson2017} used forward modeling with isotropic noise vectors to show that the resulting spin magnitude follows a noncentral-$\chi$ distribution, which he used to infer the underlying distribution.  Our $F(q)$ correction captures the corresponding increase in the RMS magnitude and allows for a condensed host-by-host description.} with past empirically-based determinations for dark matter halo spin accuracy \citep{bullock2001,Bett2007,Benson2017}.

Figure \ref{fig:finite-n-vs-n} shows an experiment to test this expectation. For every dark matter halo in each of our four simulation samples, we measure the ``true" spin $\lambda = \lambda_{\rm dm}$ using all of its particles ($N_{\rm tot} > 10^6$ in all cases).  We then draw a random subsample of $N$ particles without replacement and measure this subsampled spin, $\lambda_N$. We do this  2000 times for each halo.  The returned RMS ratios of $\lambda_N/\lambda$ are plotted as a function of $N$ in Figure \ref{fig:finite-n-vs-n}. Each simulation suite is plotted separately. Lines show medians and bands show 16th--84th percentiles. The black dashed curve is the analytic prediction given by Equation~\ref{eq:finite-n-rms} evaluated at the combined host median $\kappa=64.2$.  We see that the expectation tracks what we observe. The scatter about the line arises because there is halo-to-halo variance in $\kappa$.

A more direct test of Equation \ref{eq:finite-n-rms} is shown in the top panel of Figure \ref{fig:finite-n-collapse}. Importantly, we have evaluated $q$ directly in each of our subsampled halos, which allows us to plot the predicted ratio (Equation \ref{eq:finite-n-rms}) without any assumptions as the black dashed line. The colored curves show the median values of RMS($\lambda_N/\lambda$) at each $q$ and the shaded bands show 16th--84th percentile ranges. We see that the result collapses along the predicted curve.  This agreement suggests that the sparse sampling bias for spin measurements is well understood.

\subsection{Directional bias from small sampling}
\label{subsec:finite-n-direction}

Sparse sampling not only biases the magnitude of angular momentum we measure, but also the direction. The same noise model introduced above allows us to recover this bias, and to show that the misalignment correlates inversely with the magnitude bias discussed above.

If $\theta_N$ is the angle between the subsampled and full angular-momentum vectors then $\cos\theta_N = \boldsymbol{j}_N\cdot\boldsymbol{j}/(|\boldsymbol{j}_N|\,|\boldsymbol{j}|).$
Because $\lambda_N/\lambda = |\boldsymbol{j}_N|/|\boldsymbol{j}|$, we can write
\begin{eqnarray}
\left(\frac{\lambda_N}{\lambda}\right)\cos\theta_N
&=&\frac{|\boldsymbol{j}_N|}{|\boldsymbol{j}|}\frac{\boldsymbol{j}_N\cdot\boldsymbol{j}}{|\boldsymbol{j}_N|\,|\boldsymbol{j}|} \\
&=& \frac{(\boldsymbol{j}+\boldsymbol{\epsilon})\cdot\boldsymbol{j}}{|\boldsymbol{j}|^2}
=1+\frac{\boldsymbol{\epsilon}\cdot\boldsymbol{j}}{|\boldsymbol{j}|^2}.
\end{eqnarray}
If we take the expectation over random subsamples and remember that $\mathbb{E}[\boldsymbol{\epsilon}]=0$, we find the following remarkable identity:
\begin{equation}
    \boxed{\mathbb{E}\left[\left(\frac{\lambda_N}{\lambda}\right)\cos\theta_N\right]=1.}
    \label{eq:finite-n-exact2}
\end{equation}

\begin{figure}[t!]
\centering
\includegraphics[width=0.49\textwidth]
{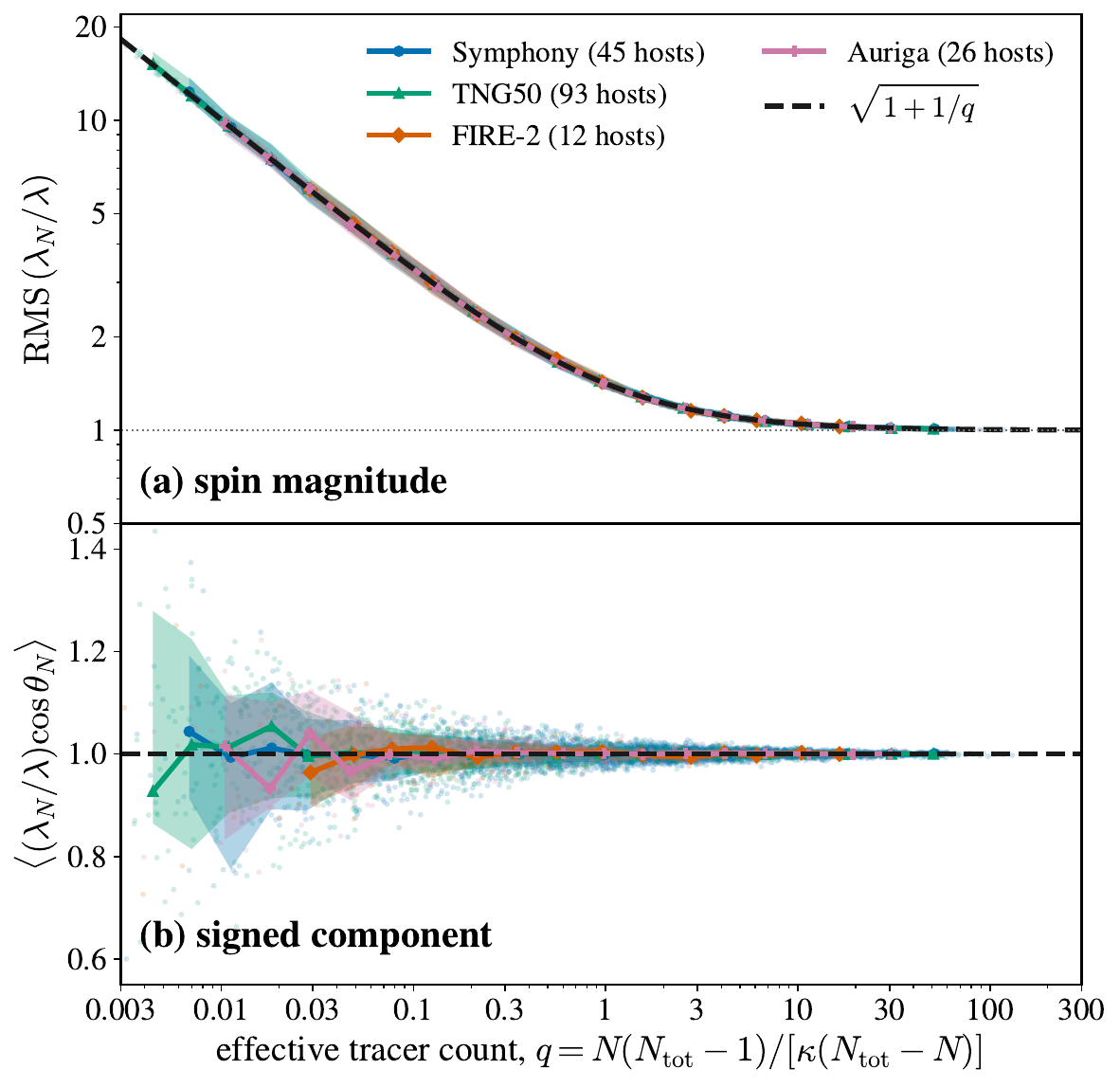}
\caption{Sparse sampling impact on measured halo spin compared to true spin (top) and on the mean signed spin component along the true angular-momentum
direction, $\langle(\lambda_N/\lambda)\cos\theta_N\rangle$ (bottom), now as a function of the tracer quotient $q$ defined in Equation~\ref{eq:finite-n-exact}.  The  black dashed curves are predictions from Equation~\ref{eq:finite-n-rms} (top) and Equation~\ref{eq:finite-n-exact2} (bottom). Faint points show individual host--$N$ measurements, with the RMS ratio (top) and mean signed component (bottom) evaluated over 2000 random subsamples. Colored lines show medians within logarithmic bins in $q$, and shaded regions span the 16th--84th percentiles of these measurements in each bin. All four suites collapse onto the same relation.}
\label{fig:finite-n-collapse}
\end{figure}

\begin{figure*}
    \centering
    \includegraphics[width=\linewidth]{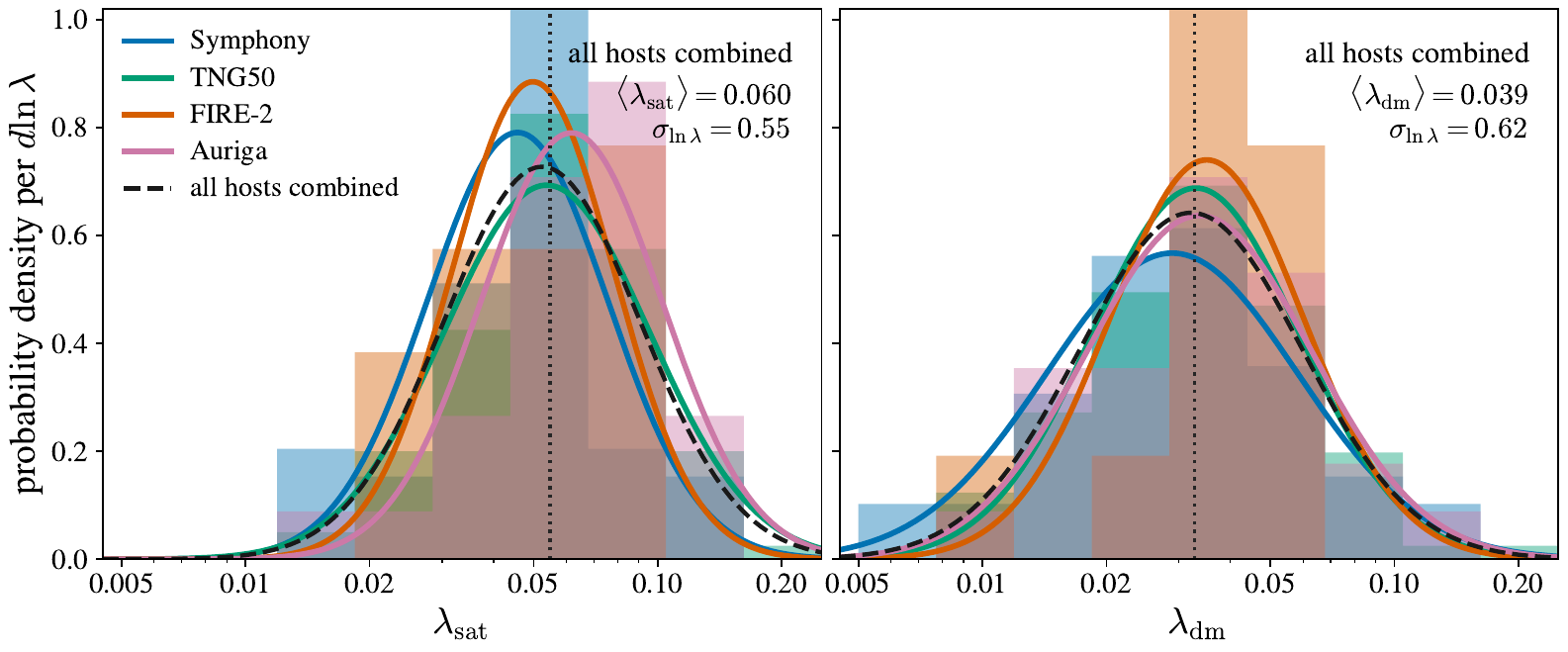}
    \caption{Spin distributions for our  fiducial samples. The left panel shows satellite system spin, $\lambda_{\rm sat}$, and the right panel shows the dark matter halo spin, $\lambda_{\rm dm}$. The colored histograms show our measured results for each suite, matching colored curves show log-normal fits to each, and the black-dashed lines show the combined sample log-normal fit. Dotted vertical lines mark the combined empirical medians, while inset values give the combined arithmetic mean and logarithmic scatter. We see that the satellite spin distributions are shifted high compared to the dark matter halo spins and have slightly less system-to-system scatter.}
    \label{fig:fiducial-spin-distributions}
\end{figure*}

The bottom panel of Figure \ref{fig:finite-n-collapse} shows that Equation \ref{eq:finite-n-exact2} is a good match to our subsampling experiments.  Here, we have used the same set of subsampled dark matter halos that we used to make the top panel to measure $\cos \theta_N$ with respect to the ``true" spin direction of the dark matter halo. As predicted, in cases with the largest magnitude bias, the directional bias is also largest in inverse proportion. 

A related point, which will be important when we explore observational tests of the satellite spin prediction in \S \ref{sec:observational}, is that when we measure the specific angular momentum along any fixed direction, there is no analogous finite-\(N\) counting bias in the corresponding dimensionless spin.
Specifically, we will consider the spin projected along the direction of disk spin. In this case, the disk-aligned component is linear in counting error:
\begin{equation}
    \widehat{\lambda}_\parallel=\widehat{\boldsymbol{\lambda}} \cdot \widehat{\boldsymbol j}_{\rm disk} = \lambda_\parallel+\frac{\boldsymbol{\epsilon} \cdot \widehat{\boldsymbol j}_{\rm disk}}{\sqrt{2}R_{\rm vir}V_{\rm vir}}.
\end{equation}
Therefore,
\begin{equation}
\boxed{
    \mathbb{E} \left[\widehat{\lambda}_\parallel\right] = \lambda_\parallel.
}
\end{equation}

\section{The Spin of Satellite Systems}
\label{sec:satspins}

With this understanding of sparse-sampling bias in hand, we now move on to measure the spin distributions of subhalo/satellite systems in our host halos.  We focus first on our fiducial satellite samples for each simulation suite. As we show below, our satellite counts per host are large enough that the expected sparse-sampling effects are small ($ \sim 10 \%$) compared to the system-to-system variance ($\sim 60\%$), which allows for a useful set of baseline predictions.  A larger sample of host halos with even more satellites per host will be needed to refine predictions on the mean to better than a $\sim 10 \%$ level.

\subsection{Measuring Point Satellite Spins}
\label{subsec:point-satellite-spins}
For each host with $N_{\rm sat}$ satellites, we measure the average specific angular momentum of the system as
\begin{equation}
 \overline{\boldsymbol j}_{\rm sat} =\frac{1}{N_{\rm sat}}\sum_{i=1}^{N_{\rm sat}}\boldsymbol j_i,
 \end{equation}
 where $\boldsymbol{j}_i =\boldsymbol{r}_i\times\boldsymbol{v}_i$ is the host-centric specific orbital angular momentum of each satellite.  This provides the associated satellite system spin for each host halo of virial radius $R_{\rm vir}$ and velocity $V_{\rm vir}$
\begin{equation}
    \lambda_{\rm sat} =\frac{|\overline{\boldsymbol j}_{\rm sat}|}{\sqrt{2}R_{\rm vir}V_{\rm vir}}.
    \label{eq:lambda-sat-definition}
\end{equation}

The analysis presented in the previous section (Equation \ref{eq:finite-n-exact}) tells us that both the number of satellites and the satellite angular-momentum contrast variable, $\kappa$, are crucial for estimating the potential for sparse-sample bias. Motivated by this, for every satellite system, we also measure the angular momentum contrast for satellites
\begin{equation}
    \kappa_{\rm sat}=\frac{\left\langle |\boldsymbol j_i|^2\right\rangle}{\left |\overline{\boldsymbol j}_{\rm sat}\right|^2}-1.
    \label{eq:kappa-sat} 
\end{equation}
The associated tracer quotient for each halo is $q_{\rm sat}= N_{\rm sat}/\kappa_{\rm sat}$.  Using Equation     \ref{eq:finite-n-rms}, this allows us to determine the predicted RMS finite-sampling enhancement on a halo-by-halo basis, $F(q_{\rm sat}) = \sqrt{1+1/q_{\rm sat}}$.

Note that the value of $\kappa_{\rm sat}$ also tells us about the within-host variance of  individual satellites’ angular-momentum {\emph{vectors}} about the satellite system’s mean vector.  If we define the dimensionless spin vector $\boldsymbol{\lambda}_i \equiv \boldsymbol j_i/\sqrt{2} R_{\rm vir} V_{\rm vir}$ then :
\begin{equation} 
    \left\langle \left|\boldsymbol{\lambda}_i-\boldsymbol{\lambda}_{\rm sat} \right|^2 \right\rangle
    = \sigma^2_{\boldsymbol{\lambda}3D} 
    = \lambda_{\rm sat}^2\kappa_{\rm sat}.
\label{eq:lambda_var}
\end{equation}
This will be relevant in our discussion of mock survey statistics in Section \ref{sec:observational}.

Table~\ref{tab:adopted-satellite-samples} summarizes these results for each simulation sample along with combined results for all samples together.  In particular, we list the median, mean, and $\ln$ scatter in $\lambda_{\rm sat}$ for each host sample. The median value of $\kappa_{\rm sat}$ is also listed for all simulations, suggesting that the angular momentum contrast for satellites is similar to that seen for dark matter particles, $\kappa_{\rm sat} \sim \kappa_{\rm dm} \sim 50$.

The ninth column in Table~\ref{tab:adopted-satellite-samples} lists the median sparse-sampling RMS enhancement factor among individual satellite systems, which is at or below $\sim 10\%$ in all cases ($F$ ranges from $1.06$ to $1.11$). For reference, this factor is much smaller than the system-to-system RMS variance $\sigma_{\ln \lambda_{\rm sat}} \simeq 0.55$, which corresponds to $\sim 60 \%$ RMS scatter and a factor of $\sim 1.7$ multiplicative or divisional scaling at one standard deviation. 

\subsection{Spin Distributions}

Figure \ref{fig:fiducial-spin-distributions} shows the  resultant satellite spin distributions (left) and dark matter halo spins (right) for each suite.   The histograms show the measured distributions for each suite. The color-matched curves show log-normal fits for each suite, and the black-dashed curves show the combined sample fits. Dotted vertical lines mark the combined empirical medians, while inset values give the combined arithmetic mean and logarithmic scatter. Both distributions have similar scatter in the log, though the satellite spin distribution is slightly narrower. Most evident is that the satellite spins tend to be higher than the dark matter spins.   We discuss the physical origin of this difference in Section \ref{subsec:whybigger}.

Tables \ref{tab:simulation-suites} and \ref{tab:adopted-satellite-samples} provide the associated median, mean, and log-normal scatter in spin for each of the suites individually and also for the combined fiducial population. Sample-wide, we find $\operatorname{med}[\lambda_{\rm sat}]=0.055$ and $\langle \lambda_{\rm sat} \rangle = 0.060$. The overall sample has a log scatter of $\sigma_{\ln} = 0.55$, which implies a linear dispersion, $\sigma_\lambda \simeq 0.036$.  Given our sample size of $H=176$, the host-counting error on the average is $\sigma_\lambda/\sqrt{H} \simeq 0.003$.

The different simulation suite medians in $\lambda_{\rm sat}$ range from $0.050$ to $0.066$.  These are similar in proportion to suite-to-suite range for $\lambda_{\rm dm}$ medians, which range from $0.030$ to $0.039$. Interestingly, the dark-matter-only Symphony suite has both the lowest $\lambda_{\rm dm}$ and lowest $\lambda_{\rm sat}$ values, but the difference compared to the hydrodynamic runs is small compared to the halo-to-halo variance in both quantities.

\begin{table*}[t!]
\centering
\normalsize
\renewcommand{\arraystretch}{1.05}
\caption{Adopted point-satellite samples and spin distributions. Listed is the number of hosts in each sample, the mean number of total subhalos per host, the median and mean of $\lambda_{\rm sat}$, the log-normal scatter (quoted in natural log), the median values of $\kappa_{\rm sat}$, $q_{\rm sat}$, and $F(q)$ (see Equation~\ref{eq:finite-n-rms}), the average cosine of the angle between satellite system spin and disk spin, the average $\lambda_\parallel$ (see Equation~\ref{eq:lambda_par}), and the average fraction of subhalo orbits that are prograde in projection with respect to the disk, $f_{\rm pro}$, averaging over all inclination angles.}
\setlength{\tabcolsep}{1.25pt}
\begin{tabular*}{\textwidth}{@{\extracolsep{\fill}}l*{11}{r}@{}}
\toprule
Suite & $N_{\rm host}$ &
\shortstack{mean\\$N_{\rm sat}$} &
\shortstack{med\\$\lambda_{\rm sat}$} &
\shortstack{mean\\$\lambda_{\rm sat}$} &
$\sigma_{\ln\lambda}$ &
\shortstack{med\\$\kappa_{\rm sat}$} &
\shortstack{med\\$q_{\rm sat}$} &
\shortstack{med\\$F(q)$} &
$\langle\cos\theta_{\rm sat,disk}\rangle$ &
$\langle\lambda_\parallel\rangle$ &
$\langle f_{\rm pro}\rangle$ \\
\midrule
Symphony & 45 & 212.6 & 0.050 & 0.051 & 0.50 & 47.1 & 4.18 & 1.11 &  &  &  \\
TNG50 & 93 & 285.9 & 0.057 & 0.062 & 0.58 & 58.7 & 4.43 & 1.11 & 0.397 & 0.0278 & 0.527 \\
FIRE-2 m12/ELVIS & 12 & 522.3 & 0.055 & 0.055 & 0.45 & 62.4 & 8.69 & 1.06 & 0.692 & 0.0383 & 0.537 \\
Auriga & 26 & 319.0 & 0.066 & 0.069 & 0.50 & 51.1 & 6.49 & 1.07 & 0.495 & 0.0427 & 0.535 \\
\midrule
Combined & 176 & 288.2 & 0.055 & 0.060 & 0.55 & 52.6 & 4.78 & 1.10 & 0.444$^{a}$ & 0.0317$^{a}$ & 0.530$^{a}$ \\
\bottomrule
\end{tabular*}
\vspace{0.4em}
\parbox{\textwidth}{\footnotesize $^{a}$ The two disk-alignment quantities and the projected prograde fraction use only the 131 hydrodynamic hosts (TNG50, FIRE-2, and Auriga). All other entries in the Combined row use all 176 hosts.}
\label{tab:adopted-satellite-samples}
\end{table*}

\subsection{Dark Matter and Satellite Spin Correlations}
\label{subsec:spin-correlations}

\begin{figure}
    \centering
    \includegraphics[width=\linewidth]{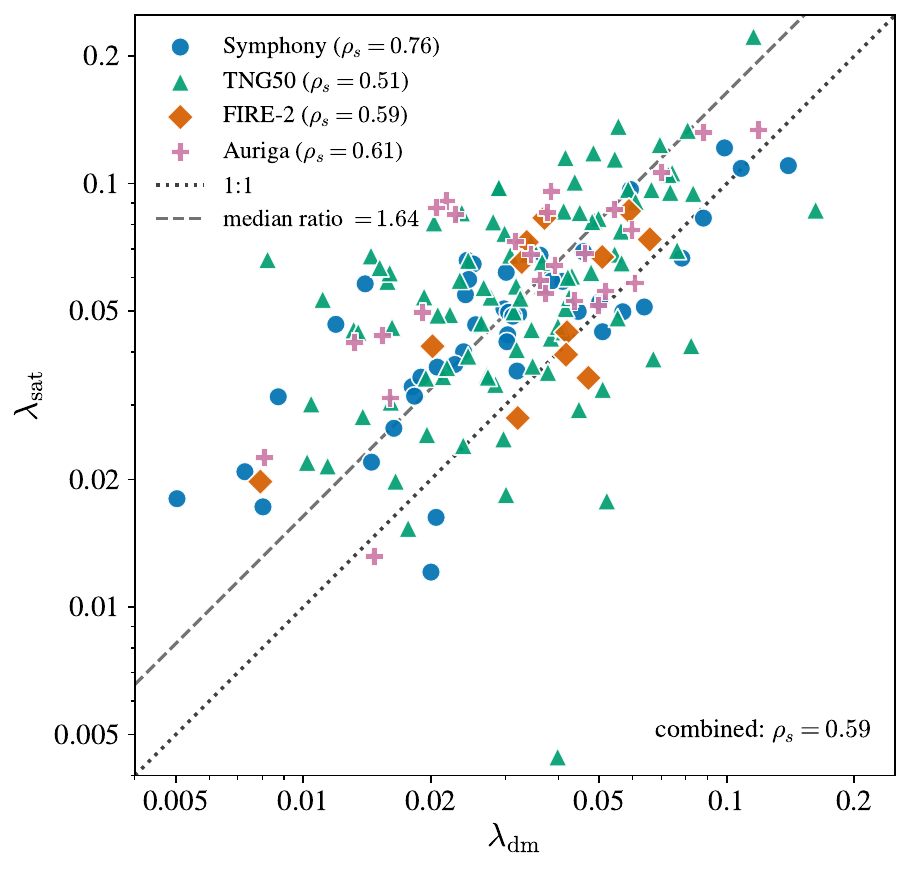}
    \caption{Host-by-host relationship between the satellite system spin, $\lambda_{\rm sat}$, and the dark matter halo spin, $\lambda_{\rm dm}$. Each point represents an individual halo, color coded by simulation suite, as indicated in the legend. The legend also shows the Spearman rank correlation, $\rho_s$, for each suite individually and for the combined sample (bottom right).  The dotted line shows the one-to-one correlation and the dashed line plots $\lambda_{\rm sat} = 1.64 \, \lambda_{\rm dm}$.} 
    \label{fig:satellite-dm-spin}
\end{figure}

Figure \ref{fig:satellite-dm-spin} plots the relationship between dark matter halo spin and satellite population spin for each host halo in our fiducial sample. Each point corresponds to a different host halo, color coded by the simulation sample, as indicated in the legend. The values are clearly correlated, though with significant scatter.  The legend also lists Spearman rank-correlation coefficients for each suite. The combined sample has a Spearman rank correlation of $\rho_s=0.59$ and a Pearson correlation of $r=0.64$, indicative of a moderately strong positive correlation, with $p<10^{-17}$. Interestingly, the dark-matter-only Symphony suite shows the most significant correlation ($\rho_s=0.76$), which may suggest that galaxy formation processes wash out the correlation somewhat. 

\subsection{Sensitivity to Subhalo Mass}
\label{subsec:mass-cut-sensitivity}
As discussed in Section \ref{subsec:satellite-selection}, our fiducial sample of satellite systems is made up of subhalos with a very low mass cut. We have done this in order to have samples with as many subhalos as possible in order to keep the counting error small. There are two potential issues with this approach.  First, for three of our four simulation samples, we are relying on halos identified with as few as $\sim 20-30$ particles, where biases in detection could render them an unfair sample of the underlying angular momentum distribution.  Second, even if these subhalos do provide a fair sample of angular momentum distributions at low mass, subhalos at larger masses (relevant for detectable satellites) may have intrinsically different angular momentum properties.

\begin{figure*}[t!]
\centering
\includegraphics[width=\linewidth]{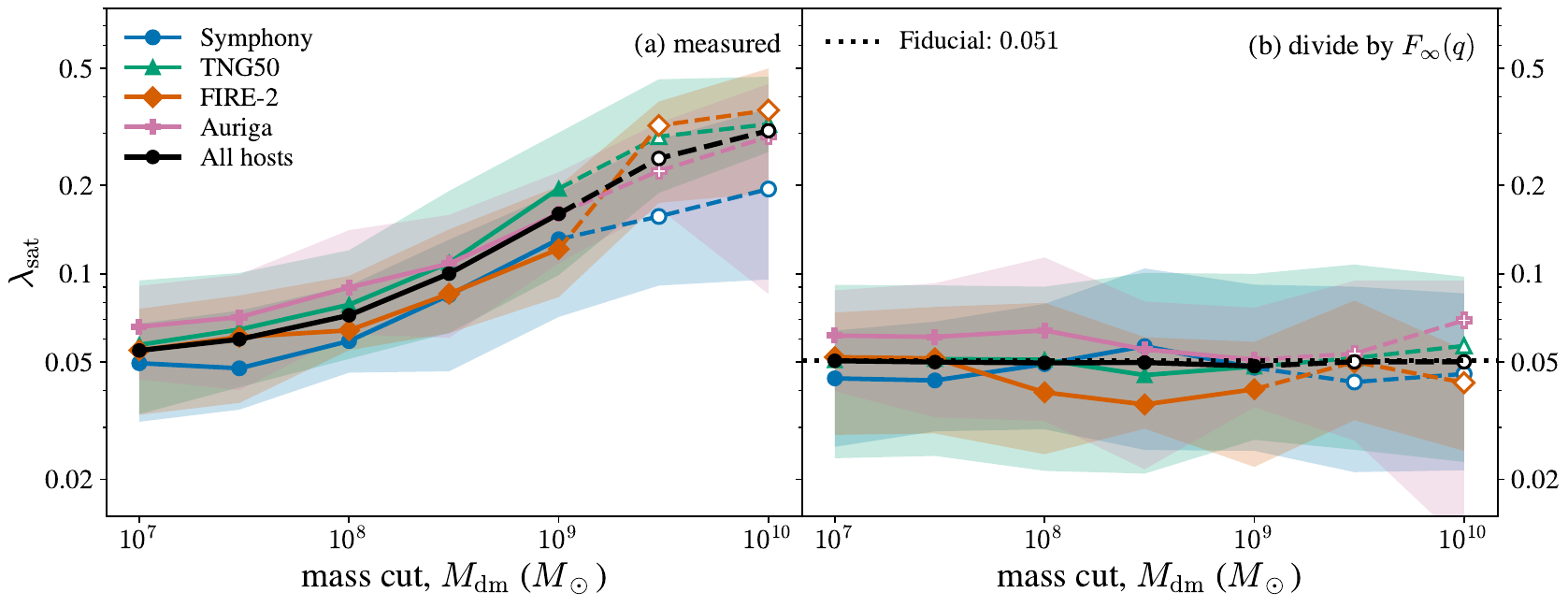}
\caption{Satellite-system spins as a function of the minimum present-day dark-matter mass: trend is driven by sparse-sampling error. Curves show host-level medians and bands span the 16th--84th percentiles. All four suites and associated host halos are included in selections from $M_{\rm dm}=10^7$ to $10^9\,M_\odot$. Above this mass, the host sample changes because an increasing fraction of the hosts contain no satellites at or above the mass cut; this is indicated by the change in line type and point type. For example, at $10^{10}\,M_\odot$, only 103 of the original 176 halos have at least one satellite.   Panel (a) shows the directly measured  $\lambda_{\rm sat}$. Panel (b) divides by the sampling enhancement in the $N_{\rm tot}\to\infty$ limit, using $q=N/\kappa_{\rm sat,parent}$, and $F(q)=\sqrt{1+1/q}$ (see text). The trend with mass becomes flat when we factor out the sampling bias.}
\label{fig:mass-ratio-invariance}
\end{figure*}

Figure \ref{fig:mass-ratio-invariance} addresses both of these questions.  The left panel plots the measured $\lambda_{\rm sat}$ distributions for host halos in each of our suites as a function of the minimum subhalo mass, $M_{\rm dm}$, used to create the sample. The points connected by lines show medians at fixed mass and the shaded bands show 16th--84th percentile ranges at each mass cut.  The black line shows the sample-wide median. We see that the measured satellite spin rises as the mass cut increases.  The qualitative direction of this increase is exactly what is expected from the sparse-counting bias discussion in Section \ref{subsec:finite-n-magnitude}. Higher mass cuts lead to fewer subhalos per host, $N$, lower $q \simeq N/\kappa$ values, and larger RMS enhancement factors.

The point types and line types change from solid to open/dashed for masses above $10^9$M$_\odot$.  This is because, above this mass, an increasingly large fraction of our host halos contain no subhalos larger than the limiting mass plotted along the horizontal axis.  That is, for $M_{\rm dm} > 10^9$M$_\odot$, the halo sample itself changes. At $10^{10}$M$_\odot$, only 103 out of the original 176 hosts have at least one satellite. 

The right panel of Figure \ref{fig:mass-ratio-invariance} scales out the expected RMS counting bias.  For each satellite halo system with $N$ satellites at every mass cut, we scale $\lambda_{\rm sat}$ downward by the RMS enhancement factor $F(q)=\sqrt{1+1/q}$.~\footnote{In general, $q=N(N_{\rm tot}-1)/[\kappa_{\rm sat}(N_{\rm tot}-N)]$, where $N_{\rm tot}$ is the total number of satellites in the parent population and $N$ is the sparse sample. Here we take $N_{\rm tot}\to\infty$, so $q=N/\kappa_{\rm sat}$, treating the fiducial catalog as a finite sample of an underlying population. We use the $\kappa_{\rm sat}$ value calculated from the fiducial ($10^7\,M_\odot$) sample. The reason $q$ decreases with mass cut is because $N$ decreases.} We see that when the RMS counting bias is scaled out, the scaled $\lambda_{\rm sat}$ distribution remains consistent with flat.~\footnote{FIRE-2 shows the most distinct trend, with a slight decline in RMS scaled spin around $\sim 10^{8.5}$ M$_\odot$. The fact that this is our smallest suite (12 halos) and none of the others show such a dip suggests that this is likely a statistical anomaly.}  The diagnostic RMS rescaling of the fiducial sample $\operatorname{med}[\lambda_{\rm sat}/F(q)]=0.051$, is plotted as a horizontal dotted line on the right. At $M_{\rm dm} > 10^{10}$M$_\odot$ the same quantity is $0.050$, which is almost identical. 

Note that dividing each halo by its \(F(q)\)  RMS counting enhancement is not an accurate way to reconstruct the underlying \(\lambda_{\rm sat}\) distribution, but it is a useful diagnostic of whether the trend with mass cut is caused by sparse sampling. The right panel should not be interpreted as a measurement of the bias-corrected median spin. The median correction will depend on the detailed shape of the underlying distribution. Nevertheless, the near removal
of the strong mass-cut dependence after this rescaling demonstrates
that the trend in the measured spin has the scaling expected from
finite sampling.  The small residual decline from approximately
$0.055$ to $0.051$ could simply represent the RMS-to-median
offset in the sparse-sampling limit. For this reason,  we will continue to quote
$\operatorname{med}[\lambda_{\rm sat}]=0.055$ as our formal result.  Higher resolution simulations or a more sophisticated treatment similar to that in \citet{Benson2017} would be required to provide a level of accuracy to better than $\sim 10\%$ on the median $\lambda_{\rm sat}$.

\subsection{Why is $\lambda_{\mathrm{sat}} > \lambda_{\mathrm{dm}}$?}
\label{subsec:whybigger}

The preceding discussion provides convincing evidence that typical $\lambda_{\rm sat}$ values are $\sim 0.05$, and thus larger than typical $\lambda_{\rm dm} \sim 0.03$ on a halo-by-halo basis. That is, this shift is not due to counting biases but rather represents a physical difference. In this section we seek to gain a simple understanding of why $\lambda_{\rm sat} > \lambda_{\rm dm}$ in most cases. 

As discussed in the introduction, the large-scale tidal field that spins up matter during halo collapse is similar for both the diffuse dark matter and the collapsed halos that are accreted as subhalos.  This suggests that any differences in the point-like spin of satellites have to do with their evolution after accretion. One obvious reason to expect a difference is that subhalos with small pericenter orbits are preferentially stripped \citep{DJM04,kelley2019,webb2020}. This naturally biases the population to have larger radii and more tangential orbits than a typical dark matter particle.

We test this by separating the spin ratio into the angular momentum carried by individual tracers and the cancellation of their vectors. From the definition of the angular-momentum contrast (Equation~\ref{eq:kappa-sat}), the net specific angular momentum of either population is $|\overline{\boldsymbol j}|=j_{\rm rms}/\sqrt{1+\kappa}$, where $j_{\rm rms}=\langle|\boldsymbol j|^2\rangle^{1/2}$. Satellites and dark matter in the same host share $R_{\rm vir}V_{\rm vir}$, so we can write their spin ratio as 

\begin{equation}
\frac{\lambda_{\rm sat}}{\lambda_{\rm dm}}=R_j  \left(\frac{1+\kappa_{\rm dm}}{1+\kappa_{\rm sat}}\right)^{1/2}, 
\qquad
R_j\equiv\frac{j_{\rm rms,sat}}{j_{\rm rms,dm}}.
\label{eq:spin_ratio_split}
\end{equation}
Here $R_j$ is the ratio of the RMS angular momentum of the individual satellites and dark-matter particles, while the second factor describes how much of that angular momentum survives cancellation in the vector sum.

\begin{figure*}[t!]
    \centering
    \includegraphics[width=\linewidth]{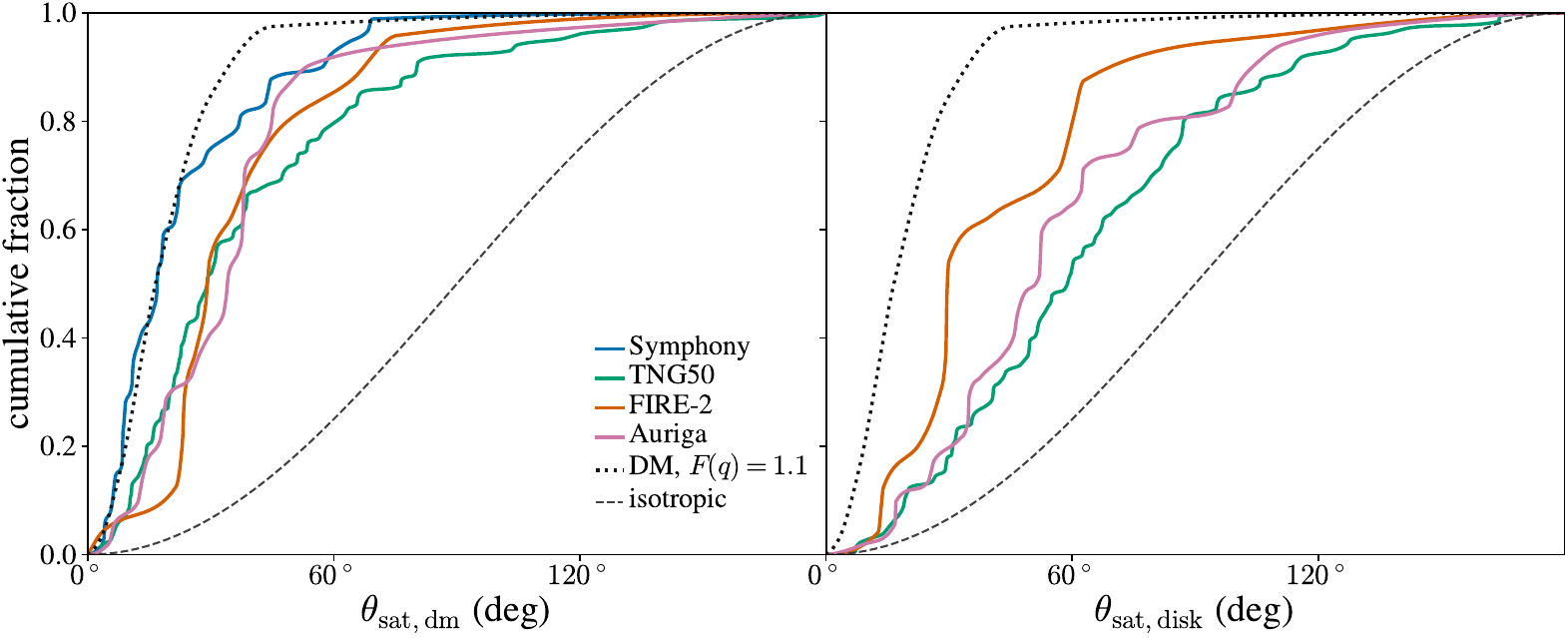}
    \caption{The cumulative distributions of angles between the point-satellite system and the host dark-matter halo spin (left) and the alignment angle between the stellar-disk angular-momentum and satellite system angular momentum (right)  for our fiducial satellite samples.  The dashed curve gives the expectation for isotropically-oriented vectors. The dotted lines show an estimate of the sparse-sample limit for this sample with $F(q) = 1.1$ (see text for details). Symphony appears only in the satellite--dark-matter comparison because it does not contain a stellar disk.}
    \label{fig:misalignment}
\end{figure*}

For our combined fiducial sample, the host-level medians and
16th--84th percentile ranges are
\begin{equation}
R_j=1.62^{+0.15}_{-0.37},
\qquad
\left(\frac{1+\kappa_{\rm dm}}{1+\kappa_{\rm sat}}\right)^{1/2}=1.07^{+0.58}_{-0.37}.
\end{equation}
We can see that $R_j>1$ drives most of the excess satellite spin, while the broader second factor accounts for most of the host-to-host scatter. 
We break $R_j$ down even further in Appendix~\ref{app:rj}, where we show that larger satellite orbital radii provide most of $R_j$, with a smaller contribution from a tendency for satellites to have more tangential orbits than dark matter particles.

\subsection{Spin Alignment}
\label{subsec:spin-alignment}
How well aligned are the angular momentum vectors of satellite systems to those of their overall dark matter halo and the central disk? We quantify this alignment using the two associated angles
\begin{equation}
 \cos\theta_{\rm sat,dm}
 =\widehat{\boldsymbol j}_{\rm sat}\!\cdot\!
  \widehat{\boldsymbol j}_{\rm dm}, \qquad
   \cos\theta_{\rm sat,disk}
 =\widehat{\boldsymbol j}_{\rm sat}\!\cdot\!
  \widehat{\boldsymbol j}_{\rm disk},
 \label{eq:satellite-alignment}
\end{equation}
where the second angle is not relevant for our dark-matter-only sample from Symphony.

Recall that each of these angles is affected by sparse sampling error.  As we derived in Equation \ref{eq:finite-n-exact2}, even systems with perfect ``intrinsic" alignment ($\cos \theta = 1$) are expected to yield measured misalignment at the level of $\cos \theta \sim  1/F(q)$ just from sparse sampling. For our fiducial samples, $F(q) \simeq 1.1$, which suggests alignments much better than $\cos \theta \sim 0.9$ should be rare, given our sample limitations.

The left panel of Figure \ref{fig:misalignment} shows the cumulative distributions of $\theta_{\rm sat,dm}$ for each of our samples as colored lines.  The isotropic expectation is shown by the dashed line.  The dotted line addresses the expected sparse-sample limit.  Here we have taken one of our Symphony dark matter halos and subsampled 410 dark matter particles over 2000 random realizations to match $F(q) = 1.1$. For each realization, we measure the cosine of the angle between the fully sampled dark matter halo and the $F(q) = 1.1$ subsample. The dotted line shows the result of this exercise.  This provides a simple estimate of the ``detection limit" of alignment angles for the characteristics of our fiducial sample of halos. 

The dark-matter-only Symphony sample shows the highest degree of alignment with the dark matter, with no systems that are counter-rotating. In fact, for small angles, the distribution tracks the sparse-sample limit quite closely.  This is perhaps not too surprising, since massive subhalos often contribute substantially to the overall angular momentum of the dark matter halo \citep{Mezini2023,Mezini2025}. However, we do see a clear signal for strongly misaligned systems, also consistent with the results of \citet{Mezini2023} and \citet{Mezini2025}.

\begin{figure*}[t!]
    \centering
    \includegraphics[width=\linewidth]{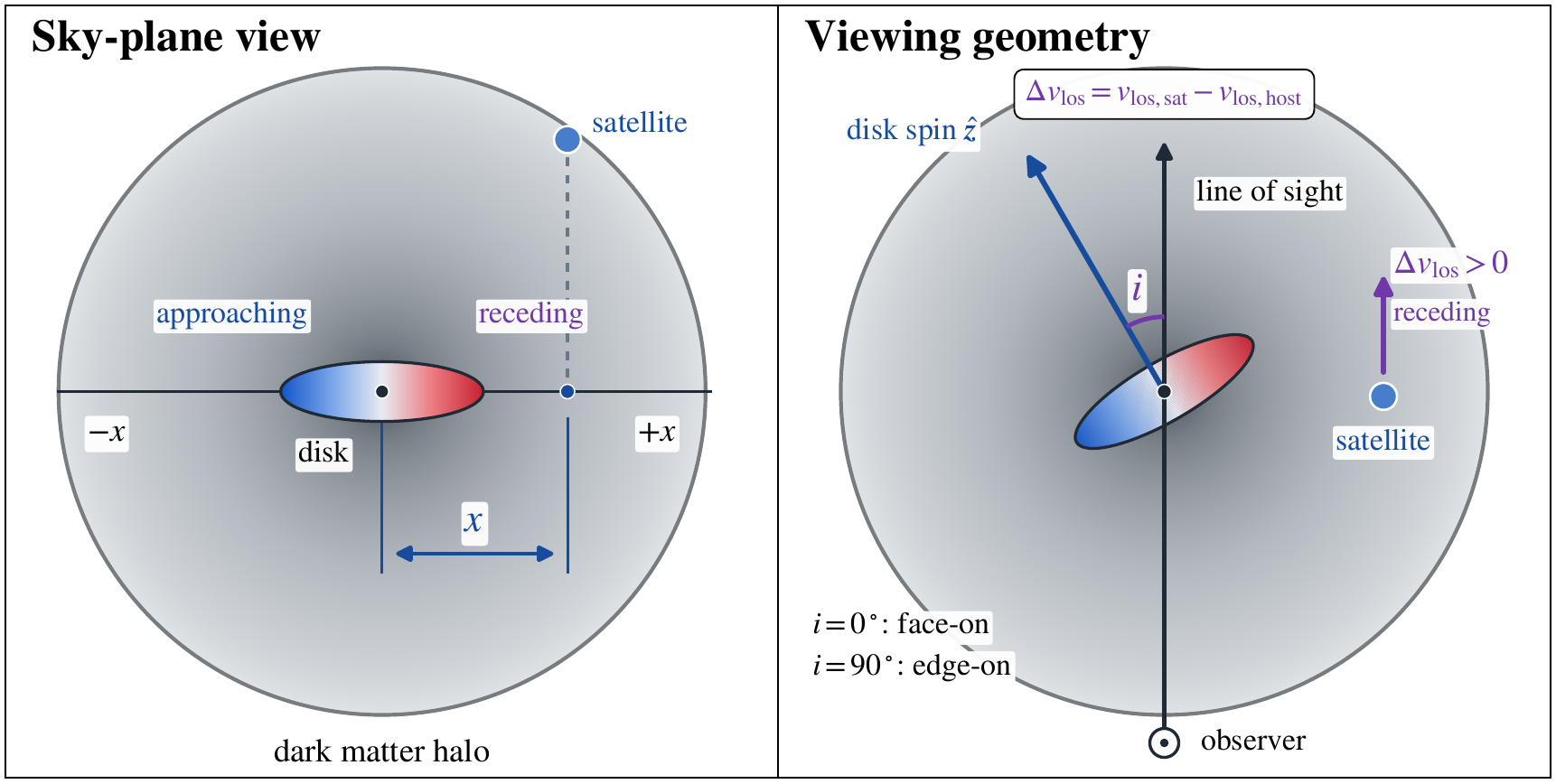}
    \caption{Observational setup and explanation of the variables in Equation \ref{eq:projected-satellite-contribution}. The cartoon on the left shows a galaxy and satellite as viewed on the sky.  The $x$ axis is set by the major axis of the disk, with positive values associated with the receding direction of rotation of the disk. The right plot illustrates the viewing geometry relative to the disk spin direction, $z$; positive line-of-sight velocities are receding. The angle $i$ defines the disk inclination with respect to the observed line-of-sight.}
    \label{fig:viewing_example}
\end{figure*}

The three suites that include galaxy formation physics in the left panel of Figure \ref{fig:misalignment} have satellite angular momentum directions that are less well aligned with the host dark matter halo, but still show significant alignment compared to isotropic. Interestingly, all three display small tails where the average satellite spin direction is more than $90^{\circ}$ off from the dark matter. The sample-wide average alignment cosine is $\langle\cos\theta_{\rm sat,dm}\rangle=0.756$, and the median misalignment angle is $\operatorname{med}(\theta_{\rm sat,dm})=26.9^\circ$. Of the 176 hosts, 168 have the two angular momentum vectors pointing in the same hemisphere.  Only eight are counter-aligned.

The right panel of Figure \ref{fig:misalignment} shows the same for  $\theta_{\rm sat,disk}$, now for the three hydrodynamic runs. As a distribution, the alignment with the disk is less than that with the dark matter halo itself, though still more aligned than would be expected for an isotropic distribution. The sample-wide median satellite--disk misalignment angle is $\operatorname{med}(\theta_{\rm sat,disk})=53.3^\circ$, while the sample-wide average misalignment angle is $\langle\theta_{\rm sat,disk}\rangle=59.6^\circ$. There is also a significant tail: $23/131=17.6\%$ ($\sim20\%$) of systems have counter-aligned net angular momenta, $\theta_{\rm sat,disk}>90^\circ$. The average values of $\cos \theta_{\rm sat,disk}$ for each of our hydrodynamic suites separately are provided in Table \ref{tab:adopted-satellite-samples}. The sample-wide average disk alignment cosine is $\langle\cos\theta_{\rm sat,disk}\rangle=0.444$.

Though not shown, we also measure the alignment angle between the dark matter halo spin and disk spin for each of our hydrodynamic runs. We find that the alignment is intermediate between what we saw for the $\cos\theta_{\rm sat,dm}$ and $\cos\theta_{\rm sat,disk}$ distributions, with $\operatorname{med}(\theta_{\rm dm,disk})=32.8^\circ$.

\section{Observables: Satellite Spins and Prograde Fraction}
\label{sec:observational}

In this section we explore the degree to which the expected spin signal from satellite systems could be observable.  We specifically explore a scenario that utilizes a set of $H$ Milky Way mass host halos, each of which has a central disk galaxy with a measured stellar mass, $M_\star$, and an observed rotation direction. Each host is surrounded by $N$ satellite galaxies with observed line-of-sight velocities, $\Delta v_{\rm los}$, and measured positions. These observables provide an estimate of the specific angular momentum of each system, and allow us to infer a dimensionless satellite spin parameter as well as measure the fraction of satellites that are orbiting in the same direction as the disk.

The primary result of this section is that observable quantities provide an unbiased estimate of the satellite system spin projected onto the angular momentum axis of the disk:
\begin{equation}
    \lambda_{\parallel} \equiv \lambda_{\rm sat} \cos(\theta_{\rm sat,disk}).
    \label{eq:lambda_par}
\end{equation}
In the limit that the satellite system angular momentum perfectly aligns with the spin of the disk, $\cos(\theta_{\rm sat,disk})=1$, then $\lambda_{\rm sat}$ can be inferred directly from observation. But more generally, as long as we have simulations that self-consistently produce disk galaxies, we can also predict $\lambda_\parallel$ directly, as we do in Section \ref{subsec:predictions} below.

Determining the spin from a measured specific angular momentum requires us to normalize by  $R_{\rm vir} V_{\rm vir} \propto M_{\rm vir}^{2/3}$.
This can be achieved, for example, using empirically derived relationships between the virial mass and observables like the galaxy stellar mass, color, rotation speed, and star formation rate  \citep[e.g.][]{ZuMandelbaum2015,Behroozi2019UniverseMachine,Cao2020}. 
For the rest of this paper, we ignore uncertainties in this mapping, but in Appendix \ref{app:halo_mass_uncertainty}, we explicitly explore the implications of reasonable uncertainties in virial mass estimates.  For example, assuming a log-normal scatter of $0.3$ dex at the Milky Way scale, we show that  the error on the mean inferred from surveys increases by $\sim 11\%$ at fixed sample size, which can be compensated for by increasing the sample size accordingly (see below).

\begin{figure*}
    \centering
    \includegraphics[width=\linewidth]{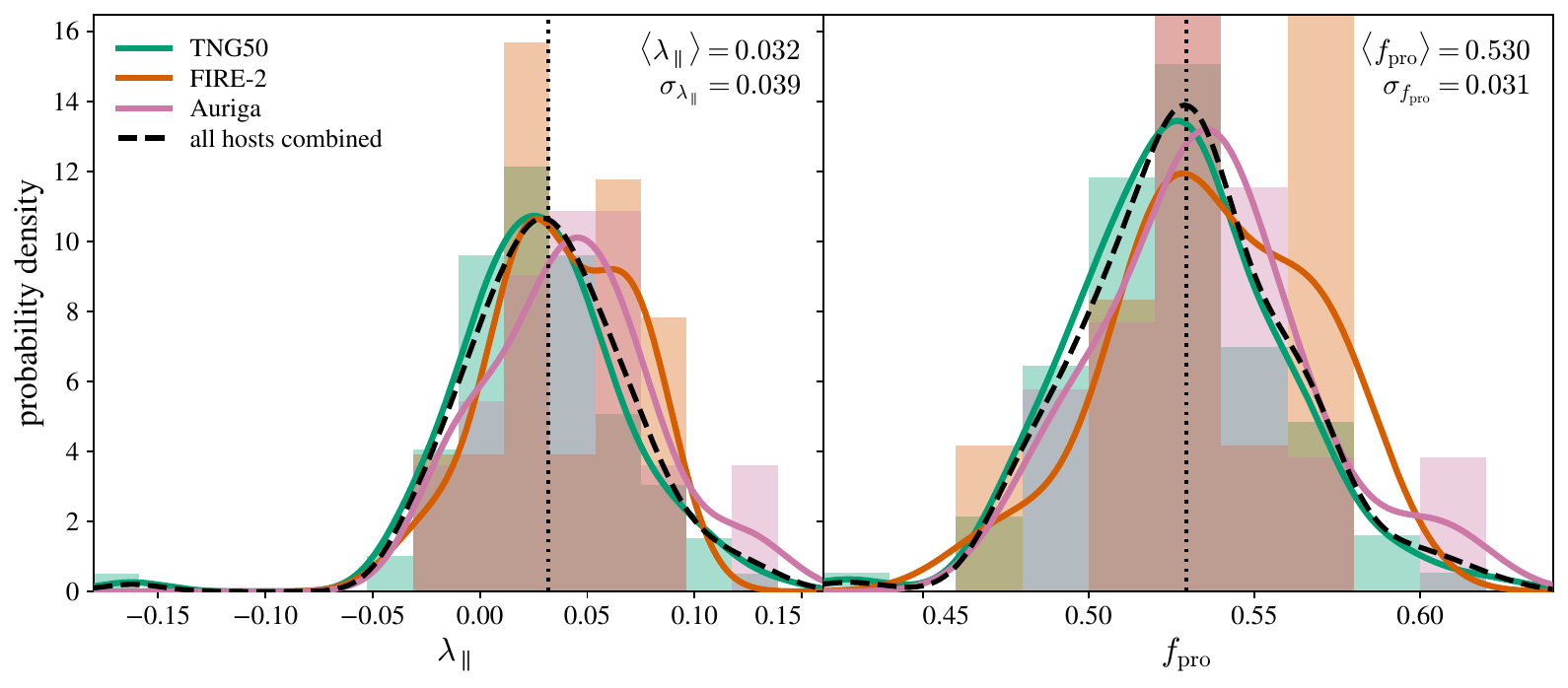}
    \caption{Idealized host-level predictions for the true signed disk-axis component $\lambda_\parallel=\lambda_{\rm sat}\cos\theta_{\rm sat,disk}$ (left) and projected, disk-referenced prograde fraction $f_{\rm pro}$ (right). The left panel includes one intrinsic three-dimensional measurement per host. The right panel shows the distribution of  host-averaged prograde fractions, each calculated by averaging over 400 isotropic sightlines. The sample consists of the 131 hydrodynamic hosts with $\langle N \rangle = 314$ satellites per host, as described in Table \ref{tab:adopted-satellite-samples}. The colored histograms and curved lines show the distributions by simulation, while the black curve is the combined sample distribution. The inset text provides the mean and host-to-host standard deviation of the combined sample.}
    \label{fig:fpro-distributions}
\end{figure*}

\subsection{Observational Setup and Analytic Expectations}
\label{subsec:observational-analytic}

Figure~\ref{fig:viewing_example} provides an illustration of our experimental setup. For each satellite, we measure its projected position $x$ along the disk major axis and its line-of-sight velocity relative to the host, $\Delta v_{{\rm los}}$. We define $x>0$ to lie on the receding side of the disk and $\Delta v_{{\rm los}}>0$ to indicate motion away from the observer. The disk inclination is $i$, where $i=0$ is face-on and $i=90^\circ$ is edge-on, and we define $z$ as the direction of disk spin. 

We show in Appendix \ref{app:projected-covariance} (Equation \ref{eq:observable}) that the expectation value for the quantity $x \Delta v_{{\rm los}}$ over satellites in the system is directly related to the $z$ component of the specific angular momentum of that system: 
\begin{equation}
    \langle x \Delta v_{{\rm los}} \rangle = \frac{\sin i}{2} \, \langle j_z \rangle.
    \label{eq:projected-moment}
\end{equation}
Recall that $z$ is the direction of spin of the {\em galaxy disk}. This means that we have an observable that is related to the satellite system angular momentum projected in the direction of disk rotation: $ j_z  = | \bar{\bf{j}}_{\rm sat}| \cos(\theta_{\rm sat,disk})$.  The factor of two appears because, for satellites drawn
uniformly in the azimuthal direction, $x \Delta v_{{\rm los}}/\sin i$ measures, on average, one of the two equal contributions to the angular momentum about the disk axis.

Based on this expectation, let us define a dimensionless variable to be measured for each satellite $s$ 
\begin{equation}
    \ell_s \equiv \frac{2x_s\Delta v_{{\rm los},s}} {\sqrt{2}R_{\rm vir}V_{\rm vir}\sin i}.
    \label{eq:projected-satellite-contribution}
\end{equation}
The values of $R_{\rm vir}$ and $V_{\rm vir}$ can be determined from the stellar mass of the central galaxy. With this definition, for a host, $h$, with $N$ observed satellites, we have a projected satellite-spin estimator 
\begin{equation}
    \widehat{\lambda}_{N,h}=\frac{1}{N}\sum_{s=1}^{N}\ell_s,
    \label{eq:projected-host-estimator}
\end{equation}
with an expectation value equal to the disk-aligned satellite spin of halo $h$:
\begin{equation}
 \mathbb{E}\!\left[ \widehat{\lambda}_{N,h}\right] = \lambda_{\parallel,h}.
 \label{eq:lambdaN_misaligned}
\end{equation}
Here $\lambda_\parallel = \lambda_z$ as defined in Equation \ref{eq:lambda_par}.

Now consider a stack of $H$ independent
hosts. For simplicity, let us assume that each host has $N$ observed satellites.  We can then construct a stacked
projected-spin estimator 
\begin{equation}
\widehat{\lambda}_{\rm stack} \equiv \frac{1}{H}\sum_{h=1}^{H}\widehat{\lambda}_{N,h}=\frac{1}{H}\sum_{h=1}^{H}\left(\frac{1}{N}\sum_{s=1}^{N}\ell_{s,h}\right).
\label{eq:stacked-spin-estimator}
\end{equation}
If the hosts are independent draws from the underlying host population, then
\begin{equation}
\label{eq:lambdastack_misaligned}
\boxed{
\mathbb{E}\left[\widehat{\lambda}_{\rm stack}\right] =\langle\lambda_{\parallel} \rangle.
}
\end{equation}
Importantly, the average of the disk-aligned spin, $\lambda_{\parallel}$, is a predicted quantity (see Section \ref{subsec:predictions}).  

Of course, our ability to measure this quantity with precision depends on our expectations for the variance of a given experimental setup. The associated variance will have two terms.  The first is the underlying host-to-host variance in $\lambda_\parallel$, which we write as $\sigma_{\lambda_\parallel}^2$.  The second is the variance in inferred spin for each host halo, $\sigma^2_{\lambda_{N,h}}$. For $H$ hosts, the total variance in the stacked estimator is
\begin{equation}
\sigma^2_{\langle \lambda_\parallel \rangle} = \frac{1}{H} \left[ \sigma_{\lambda_{\parallel}}^{2} + \langle \sigma^2_{\lambda_{N,h}}\rangle_h \right].
\label{eq:variance_stack1}
\end{equation}

The first term, the host-to-host variance, is something we can measure directly in our simulations, and we find $\sigma_{\lambda_\parallel}^2 \sim \lambda_{\parallel}^2$ (see Figure \ref{fig:fpro-distributions} below). The second term comes from the intrinsic variance in
\(\ell_s\) among the satellites of a given host $h$.  It is proportional to the expected variance in individual satellites' angular momentum vectors about the system mean  ($\lambda_{\rm sat}^2 \, \kappa_{\rm sat}$, Equation \ref{eq:lambda_var}) modulo a geometric factor that depends on the inclination angle:
\begin{equation}
\sigma_{\ell,h}^{2}
=
C_{\rm proj}
\lambda_{{\rm sat},h}^{2}\kappa_{{\rm sat},h}.
\end{equation}
We show in Appendix \ref{app:projected-covariance} (Equation \ref{eq:sigma_ell_final})
that under reasonable assumptions, $C_{\rm proj} = 2/ (3\sin^2i)$. 
For $N$ satellites, the sampling variance in $\widehat{\lambda}_{N,h}$ becomes $ \sigma^2_{\lambda_{N,h}}\ = \sigma^2_{\rm \ell, h}/N$.  This allows us to write the total variance in the
stacked estimator 
\begin{equation}
\sigma^2_{\langle \lambda_\parallel \rangle} = \frac{1}{H} \left[ \sigma_{\lambda_\parallel}^{2} + \frac{\left\langle \sigma_{\ell,h}^{2}\right\rangle_h}{N} \right].
\label{eq:variance_stack}
\end{equation}
For $N \lesssim 100$, the second term dominates and the dispersion is~\footnote{If $N$ varies with host, then $\sqrt{1/N} \rightarrow \sqrt{\langle 1/N\rangle}$ in Equation \ref{eq:error_on_mean}.}
\begin{equation}
    \sigma_{\langle\lambda_\parallel\rangle}  \simeq \sqrt{\frac{ \lambda_{\parallel}^2 \kappa_{\parallel}}{HN}}\simeq \frac{0.4}{\sqrt{HN}},
    \label{eq:error_on_mean}
\end{equation}
where~\footnote{$\kappa_\parallel =
\langle\sigma_{\ell,h}^2\rangle_h/\langle\lambda_\parallel\rangle_h^2.$} $\kappa_{\parallel} \equiv C_{\rm proj} \lambda_{\rm sat}^2\kappa_{\rm sat}/\lambda_{\parallel}^2$.
In the last step we assume a random survey restricted to high inclinations, $i > 45^\circ$, which gives
 $C_{\rm proj} \simeq 2/3 \, \langle 1/\sin^2i\rangle \simeq 0.83$ and $\kappa_\parallel \simeq 160$. The implication is that a sample with the total number of observed satellites obeying $H N \gg \kappa_{\parallel} \sim 160$ will allow a precise measure of $\langle \lambda_\parallel \rangle$. For example, $H = 500$ hosts with $N=4$ satellites each would yield a relative error on the mean of $\sim 30\%$.

Another observable quantity related to satellite spin is $f_{\rm pro}$, the fraction of satellites that are orbiting in the same direction as the stellar disk. A satellite is classified as prograde when its projected position and line-of-sight velocity have the same sign: $x_s\Delta v_{{\rm los},s}>0$. In Appendix \ref{App:co-rot} we show that the error on the mean of $f_{\rm pro}$ inferred from our sample setup is well approximated by~\footnote{If $N$ varies with host, then $\sqrt{1/N} \rightarrow \sqrt{\langle 1/N\rangle}$ in Equation \ref{eq:error_on_f}.}
\begin{equation}
    \sigma_{\langle f_{\rm pro}\rangle}
    \simeq \frac{1}{2\sqrt{HN}}.
    \label{eq:error_on_f}
\end{equation}
As with spin mean errors in Equation \ref{eq:error_on_mean}, samples with large $H \times N$ will allow us to measure the projected prograde fraction with a good degree of precision. 

\subsection{Predicted Observables}
\label{subsec:predictions}

In this subsection we use our fiducial set of satellite catalogs taken from our three hydrodynamic suites to provide idealized predictions for our observables $\lambda_{\parallel}$ and $f_{\rm pro}$. This sample includes 131 hydrodynamic hosts with $\langle N \rangle = 314$ satellites per host. Of course, satellites this small will be largely unobservable -- we use them only to provide an accurate prediction for a parent sample.  Section \ref{sec:mock-surveys} is devoted to more realistic mock-observing scenarios with many fewer satellites and direct tests of our analytic expectations.

\begin{figure*}
    \centering
    \includegraphics[width=\linewidth]{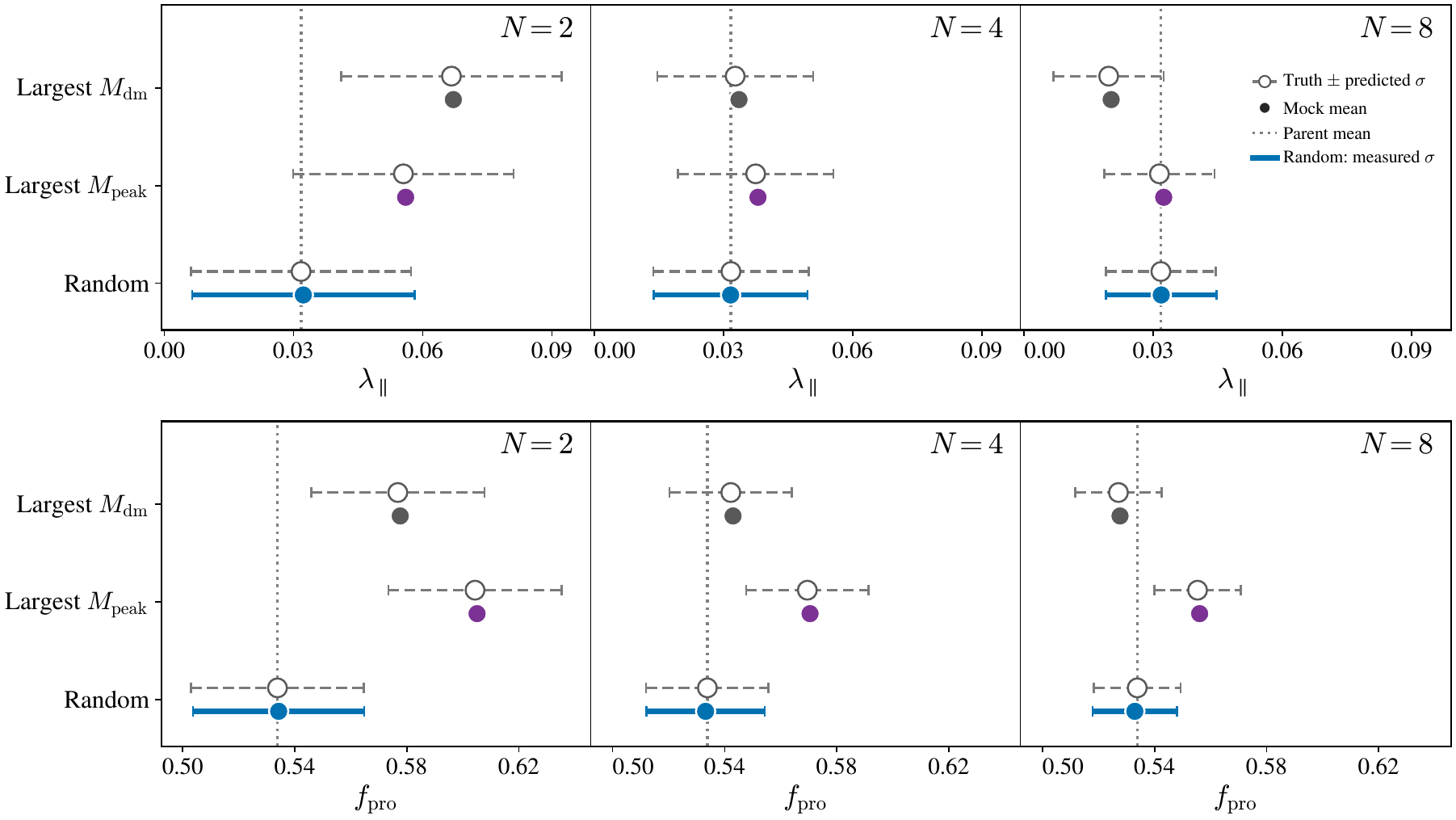}
    \caption{Mock survey results for inferred projected spin (top) and projected prograde fraction (bottom), using $N=2,4,8$ satellites per host (columns) and $H=131$ hosts per survey. Gray, purple, and blue circles select the largest present-day dark-matter masses, largest peak halo masses, and random subhalos, respectively. Solid circles show observed average values from each relevant mock sample, while open circles show the selected-sample truths. In every case, the mock means match closely to the truth, demonstrating that the estimator is unbiased. The dashed gray error bars show the predicted sample variance for a given $N$ according to Equations \ref{eq:error_on_mean} and \ref{eq:error_on_f}. Dotted lines mark the fiducial-sample means; the prograde reference matches the mocks, retaining only $i\geq45^\circ$.}
    \label{fig:direct-survey-predictions}
\end{figure*}

The left and right panels of Figure \ref{fig:fpro-distributions} show the predicted distribution of our observables, $\lambda_{\parallel}$ and $f_{\rm pro}$, respectively, for our fiducial satellite samples.  For  $\lambda_\parallel$, we use one intrinsic 3D measurement per host, resulting in a distribution of 131 values. For $f_{\rm pro}$, we calculate the projected fraction along 400 isotropic sightlines without an inclination cut, then average these fractions within each host before constructing the distribution. Thus, the distribution represents 131 line-of-sight-averaged values, with one per host.  The colors designate the different suites, as indicated.   Note that unlike $\lambda_{\rm sat}$, the disk-aligned satellite spin can be negative, corresponding to cases when the satellites are orbiting in the opposite direction of the disk.  
The black dashed lines are the combined host distributions and the vertical dotted lines mark their means. 

The sample-wide averages $\langle \lambda_\parallel \rangle$ and $\langle f_{\rm pro} \rangle$, along with the measured dispersions, $\sigma_{\lambda_\parallel}$ and $\sigma_{f_{\rm pro}}$, are listed in their respective panels.  For our host sample size of $H=131$, we estimate the measured error on their means as $\sigma/\sqrt{H}$ and write our formal predictions as $\langle \lambda_\parallel \rangle = 0.032 \pm 0.003$ and $\langle f_{\rm pro} \rangle = 0.530 \pm 0.003$. Table \ref{tab:adopted-satellite-samples} lists the values of $\langle \lambda_{\parallel} \rangle$ and $\langle f_{\rm pro} \rangle$ for each of the relevant suites, along with the sample-wide results. We note that $\langle \lambda_{\parallel}\rangle/\langle \lambda_{\rm sat}\rangle \simeq 0.5$ in our hydrodynamic simulations. This value is slightly larger than $\langle \cos \theta_{\rm sat, disk} \rangle = 0.44$, which is consistent with a modest positive covariance between $\lambda_{\rm sat}$ and $\cos \theta_{\rm sat, disk}$.

It is important to note that while the intrinsic $\lambda_\parallel$ values do not depend on viewing inclination, $\langle f_{\rm pro} \rangle$ does. The quoted value of $\langle f_{\rm pro} \rangle = 0.530 \pm 0.003$ averages over all inclinations, but if we restricted to sightlines with $i\geq45^\circ$ instead, then $\langle f_{\rm pro}\rangle=0.534$. A more detailed exploration on how $f_{\rm pro}$ changes with the viewing angle can be found in Appendix~\ref{sec:inclination}.

We find that the true fraction of individual satellites, averaged over hosts, that are rotating in the same direction as the disk in 3D has an average value $\langle f_{\rm pro}^{\rm 3D} \rangle = 0.554$.  As discussed in Appendix \ref{sec:inclination}, the mapping between $f_{\rm pro}^{\rm 3D}$ and the observable projected prograde fraction $f_{\rm pro}$ tracks a well-defined trend with disk inclination.  
In particular, we find that the observed prograde fraction follows $f_{\rm pro} \simeq 0.5 + 0.037\, \sin i$.  This expectation will be important to keep in mind when comparing to observational measurements in detail.

\section{Testing Methodology}
\label{sec:mock-surveys}
This section is devoted to testing our ability to measure 
$\langle \lambda_\parallel \rangle$ and $\langle f_{\rm pro} \rangle$ using the estimators discussed in the previous section (Equation \ref{eq:lambdastack_misaligned}) as well as our expectations for sample variance ($\propto 1/HN$; see Equations \ref{eq:error_on_mean} and \ref{eq:error_on_f}).   We rely on mock surveys, with geometry depicted in Figure \ref{fig:viewing_example}. In what follows, we restrict to sight lines that produce disk inclinations $i >45^\circ$. This is motivated by the fact that the accuracy in determining $\lambda_\parallel$ is expected to scale as $\propto 1/\sin^2 i$ (Equation \ref{eq:error_on_mean}).~\footnote{This inclination cut is familiar to those often used in rotation curve studies since minor variations in $i$ can have large implications for the implied velocity at small inclination ($j \propto 1/\sin i$).} Of course, future surveys could explore scenarios that weighted measurements appropriately by the disk inclination in order to maximize the information collected. 

Note that in order to construct the $\lambda_\parallel$ estimator,  we must divide measured quantities by $R_{\rm vir}$ and $V_{\rm vir}$ (Equation \ref{eq:projected-satellite-contribution}). In what follows, we adopt an idealized scenario where we know each host's virial quantities perfectly and adopt the true host $R_{\rm vir}$ and $V_{\rm vir}$ in each case. Any real survey would rely on estimates for these quantities. In Appendix \ref{app:halo_mass_uncertainty} we discuss how standard empirical estimates  would likely increase the variance by modest factors $\sim 24\%$, which could be compensated for by increasing the sample size ($HN \rightarrow \, 1.24 \, HN$).

Finally, in a real observational survey, we would apply projected position and velocity cuts to select satellites and limit contamination. In the idealized mocks presented below, we observe all satellites within each host's $R_{\rm vir}$ without a velocity cut.  While restricting to subhalos known to be within the virial radius, we tested scenarios with radial cuts $10\leq R_{\rm proj}\leq300\,{\rm kpc}$, $10\leq R_{\rm proj}\leq283\,{\rm kpc}$, and velocity limits of $275$, $400$, and $500\,{\rm km\,s^{-1}}$.  We find no significant differences to the results presented below.  The inferred $\lambda_\parallel$ changes by less than $10\%$ and the mean $f_{\rm pro}$ changes by less than $1\%$. Followup work utilizing large boxes to explore contamination will be needed to design best strategies to remove contaminants.

In the first set of mocks,  we consider samples with $N=2$, $4$, or $8$ satellites chosen using three different methods of satellite selection: the $N$ largest present-day dark matter masses ($M_{\rm dm}$), the $N$ largest peak halo masses ($M_{\rm peak}$), and a random selection of $N$ subhalos out of our full subhalo sample. Here, $M_{\rm peak}$ is the maximum total bound subhalo mass along its main-progenitor history. 

We perform a mock survey as follows. For each host, $h$, we generate 400 isotropic sight lines and keep only those with a disk orientation $i\geq45^\circ$. We then draw one of the retained sight lines at random and use its $N$ identified satellites to calculate $\lambda_{\parallel, h}$ and $f_{{\rm pro},h}$.  We do this for all $H=131$ hosts and calculate the stacked average of $\lambda_\parallel$ and $f_{\rm pro}$ for this particular mock survey. Finally, we perform the above set of steps 1800 times, which produces a distribution of measured averages, including the many-survey average of averages. 

Figure~\ref{fig:direct-survey-predictions} summarizes the results of this exercise, with $\lambda_\parallel$ measurements presented in the top set of boxes and $f_{\rm pro}$ along the bottom.  The panels from left to right present results for the $N=2, 4$, and $8$ samples; the vertical axis denotes the method for choosing the $N$.  The vertical dotted line in each panel is the full parent population mean of each relevant quantity.  Note that the parent mean in $f_{\rm pro}$ plotted is calculated for the same orientation cut used in the mock $i \ge 45^\circ$ ($0.534$), which is slightly higher than the isotropic result ($0.530)$.

The average of each set of mocks for either measured quantity is shown by the filled circles, color coded by the selection criteria. These should be compared to the open gray circles, which represent the true underlying values of samples measured. The error bars attached to the open circles are the predicted $\pm 1\sigma$ dispersion for the relevant $HN$ value as given by Equations \ref{eq:error_on_mean} and \ref{eq:error_on_f}.

The first, and most important result, is that the average of the estimator (colored circles) is very close to the true population average (open circles) in every case. 
For the random-N draws (blue), in particular, the true average (as well as the measured average) is indistinguishable from the full parent population average.  This is to be expected, since, by randomly drawing $N$ subhalos many times, we effectively sample the full population. The random-N experiments also allow us to test whether the predicted dispersions match our expectations.  The blue error bars in each panel show the measured dispersion for the relevant mock surveys.  In every case, both for $\lambda_\parallel$ and $f_{\rm pro}$, the random $N$ dispersion is within $\sim 10\%$ of the predicted dispersion.  Given the approximations we used in deriving the expected dispersion, this is excellent agreement.

\begin{figure}
\centering
\includegraphics[width=\linewidth]{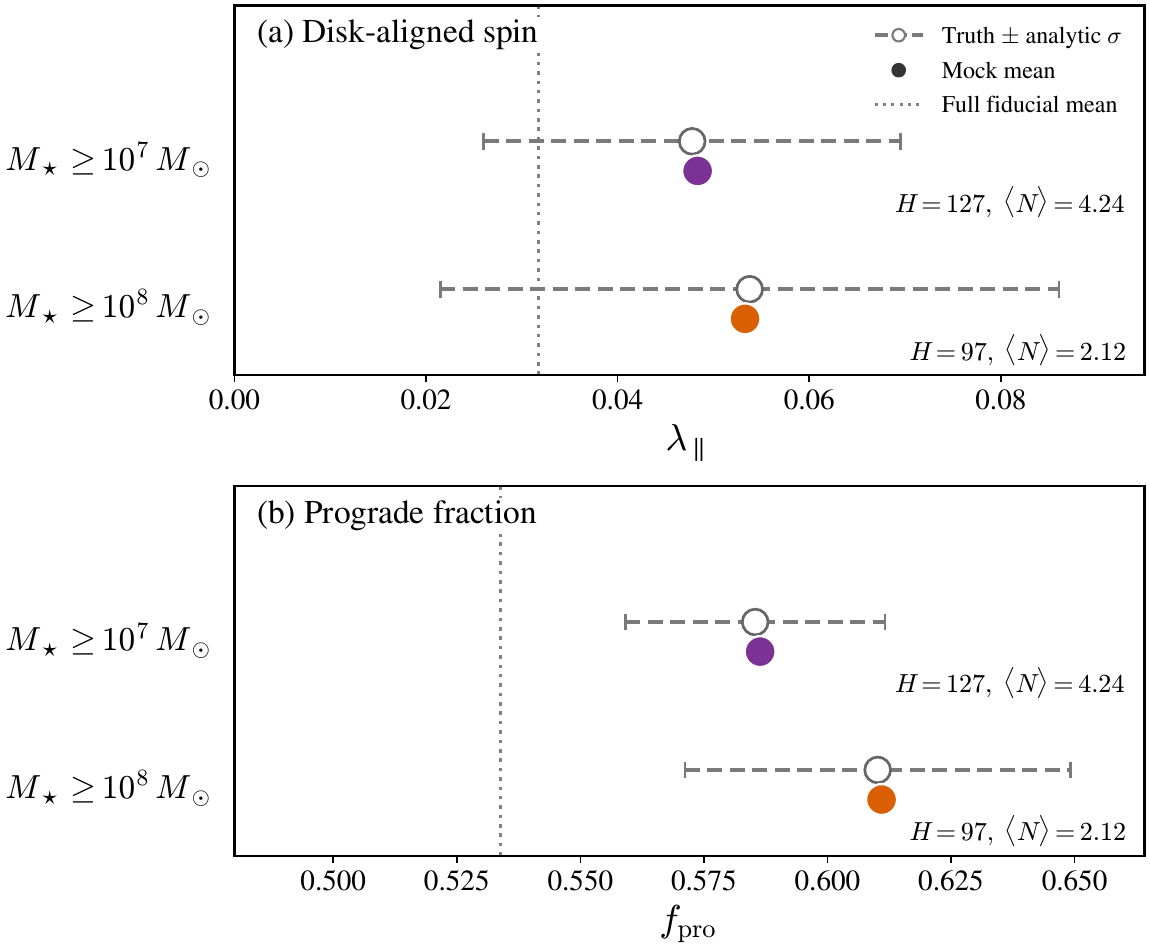}
    \caption{Predicted mean disk-axis spin (top) and projected prograde fraction (bottom) for satellites selected by stellar mass. Purple and orange filled circles show means of multiple mocks that include satellites that have $M_\star\geq10^7\,M_\odot$ and $M_\star\geq10^8\,M_\odot$, respectively. Each mock includes every eligible host exactly once, giving $H=127$ and $97$, respectively. Open circles and dashed gray bars show the selected-sample truth and the expected counting dispersion for samples of this size from Equations~\ref{eq:error_on_mean} and~\ref{eq:error_on_f}. Insets show the mean satellite count for each sample selection. Dotted lines mark the full fiducial-sample means.}
\label{fig:stellar-mass-cuts}
\end{figure}

The $M_{\rm dm}$ and $M_{\rm peak}$ selections for the largest $N$ satellites are qualitatively different than the random $N$ choice.  In these cases, we are always observing the same $N$ satellites.  The multiple surveys are constructed by simply viewing their hosts from different angles. In this circumstance, it is reasonable to expect that the sample means will be different than the overall parent mean, and indeed they are. The ranked selections shift the means relative to the parent population, with the direction and magnitude of the shifts
depending on $N$. Both ranked selections give higher means than the parent population for $N=2$, but the $M_{\rm dm}$ selection has a lower mean than the population average for $N=8$.

It is interesting to ask whether the difference between the sub-sampled mean from massive satellites and the full parent mean is consistent with a variation from counting statistics or a true physical difference associated with the fact that they are the most massive satellites. Although we do {\em not} see any evidence for a physical increase in the  3D spin $\lambda_{\rm sat}$ as we look at higher mass subhalos (See Figure \ref{fig:mass-ratio-invariance}), it is possible, for example, that the physical correlation between the disk spin direction and satellite spin direction is stronger for the most massive subhalos. 

The degree to which the differences should be expected from counting statistics alone can be gauged by looking at the dashed error bars, which show the predicted counting dispersion from Equations \ref{eq:error_on_mean} and \ref{eq:error_on_f}. For the $\lambda_\parallel$ measurements, the $M_{\rm peak}$ choices are all within $1 \sigma$ of the parent mean, and the $M_{\rm dm}$ choices are all within $1.5 \sigma$.  The projected prograde fractions show larger offsets. For $N=2$, the $M_{\rm peak}$ selection is $2.2 \sigma$ high and the $M_{\rm dm}$ choice is $1.4 \sigma$ high. The larger $N$ cases are less significantly different from the parent means, especially considering that these samples are both inclusive of the $N=2$ samples. While it is possible that there is a true, physical difference in $f_{\rm pro}$ for the most massive subhalos, our limited sample size does not allow us to determine whether what we see for the $N=2$ prograde fraction is anything other than a $\sim 2\sigma$ statistical fluctuation.

We cannot use the $M_{\rm peak}$ and $M_{\rm dm}$ selection mocks to test the variance predicted in Equations \ref{eq:error_on_mean} and \ref{eq:error_on_f}.  This is because these mocks only include the variance associated with viewing orientation and rely on the same $N$ satellites every time. However, the orientation variance can be estimated (Equation \ref{eq:orientation_variance}) and, though not shown, we find that the observed scatter in our mocks is consistent with that estimate.

Although the results presented in Figure~\ref{fig:direct-survey-predictions} are useful for testing our methodology, a real survey will be limited to some fixed luminosity or stellar mass threshold. Figure \ref{fig:stellar-mass-cuts} presents such an example. We have performed the same mock strategy described above, except now applying uniform cuts of $M_\star\geq10^7\,M_\odot$ and $M_\star\geq10^8\,M_\odot$, keeping all satellites above the cut to calculate $\lambda_\parallel$ and $f_{\rm pro}$. 
Hosts with no qualifying satellites are excluded in each case, leaving 127 and 97 hosts, with mean satellite counts of 4.24 and 2.12, respectively. As in Figure~\ref{fig:direct-survey-predictions}, filled colored circles show the average values over all mocks, and these match well the true values for the underlying samples, shown by the open circles. The dashed error bars show the estimated dispersion given the sample size, with  $1/N$ replaced by $\langle 1/N \rangle$ in Equations \ref{eq:error_on_mean} and \ref{eq:error_on_f}. 

For $\lambda_\parallel$, both stellar-mass-selected sample means are less than $1\sigma$ from the full population expectation. 
For $f_{\rm pro}$, both stellar-mass cuts have sample means that are approximately $2\sigma$ above the plotted expectation. Examining the non-overlapping interval $10^7\leq M_\star/M_\odot<10^8$ separately gives a projected prograde fraction that is approximately $1\sigma$ above the parent sample mean. These results are similar to what we found for the $M_{\rm peak}$ selections in Figure~\ref{fig:direct-survey-predictions}, which is not surprising since there is significant overlap (that is, high $M_{\rm peak}$ correlates well with high $M_\star$). As emphasized above, our limited sample size does not allow us to determine whether the high $f_{\rm pro}$ value of this subsample is anything other than a $\sim 2\sigma$ statistical fluke. Confirming the observed offset at $3 \sigma$ for satellites of this stellar mass would require a sample of simulated hosts that is $\sim 2.5$ times larger than what we are using in this paper.  
 The DREAMS suite, for example, with its $1024$ Milky-Way-mass zooms, could provide an ideal means to look for such an effect \citep{Rose2025DREAMS,Rose2026DREAMSSatellites}.

\section{Conclusions}
\label{sec:Conclusions}

We have studied the collective orbital angular momentum of subhalo/satellite systems around Milky Way size hosts using a dimensionless spin parameter, $\lambda_{\rm sat}$, in analogy to the spin parameter used to characterize the angular momentum of dark matter halos. We also explored ways to observe a related quantity, $\lambda_\parallel$, the orbital spin projected along the direction of its host disk axis. Our primary results are as follows:

\begin{itemize}
    \item  Using 176 Milky Way mass hosts, taken from both dark matter (Symphony) and hydrodynamic simulations (TNG50, FIRE-2, Auriga), we measured a median satellite spin of $\lambda_{\rm sat}=0.055$, with a log-normal scatter of $\sigma_{\rm ln} = 0.55$ (Figure \ref{fig:fiducial-spin-distributions},  Table \ref{tab:adopted-satellite-samples}). 

    \item Halo-by-halo, the satellite system spin, $\lambda_{\rm sat}$, correlates with the dark matter halo spin, $\lambda_{\rm dm}$, with a median ratio $\lambda_{\rm sat}/\lambda_{\rm dm} \simeq 1.64$ (Figure \ref{fig:satellite-dm-spin}).
    
    \item Using sampling theory, we derived closed-form expressions for the relationship between tracer counts and spin measurement bias (Equations \ref{eq:finite-n-exact} and \ref{eq:finite-n-exact2}); accounting for this, we find no evidence that $\lambda_{\rm sat}$ has a dependence on subhalo mass threshold (Figure \ref{fig:mass-ratio-invariance}).

    \item Using 131 Milky Way-like disk hosts from the hydrodynamic simulations, we showed that satellite system spin directions are preferentially aligned with disk spin directions, with an average alignment $\langle \cos(\theta_{\rm sat, disk})\rangle = 0.444$ (Figure \ref{fig:misalignment} and Table \ref{tab:adopted-satellite-samples}).

    \item We showed analytically (Equations \ref{eq:projected-moment} -- \ref{eq:lambdastack_misaligned}) and with mock surveys (Figures \ref{fig:direct-survey-predictions} and \ref{fig:stellar-mass-cuts}) that it is possible to measure the average satellite spin projected onto the disk rotation axis, $\lambda_\parallel = \lambda_{\rm sat} \cos(\theta_{\rm sat, disk})$, accurately in observations by “stacking” galaxy disk satellite systems and observing 1) the sense of disk rotation, 2) projected satellite positions, and 3) projected satellite velocities (See Figure \ref{fig:viewing_example}). 
    
    \item Using our sample of 131 hydrodynamic disk galaxy simulations, \textbf{we predict this observable spin parameter to have an average value $\langle\lambda_\parallel\rangle = 0.032$} (Figure \ref{fig:fpro-distributions}).  Coincidentally, this observable expectation is numerically close to the underlying average 3D dark halo spin $ \langle \lambda_\parallel \rangle \simeq \langle \lambda_{\rm dm} \rangle$. We also predict a related observable: \textbf{the fraction of satellite galaxies that should be observed to have orbits that are prograde with respect to the host disk spin in projection: $\langle f_{\rm pro}\rangle\simeq0.53$} (Figure \ref{fig:fpro-distributions} and Appendix \ref{sec:inclination}).

    \item Using analytic approximations (Equations \ref{eq:error_on_mean} and \ref{eq:error_on_f}) and  mock surveys (Figure \ref{fig:direct-survey-predictions}), we showed that, for practical applications, the precision of an experiment to measure $\langle\lambda_\parallel\rangle$ and $f_{\rm pro}$ from a stacked survey is inversely proportional to the number of host disks $H$ and number of satellites per host, $N$, such that $\sigma^2 \propto 1/HN$. 

\end{itemize}

Although the spin of satellite systems has not been investigated directly in observations, several related efforts exist.  \citet{Zaritsky1997} performed a stacking experiment using 115 satellite galaxies orbiting 69 isolated spiral galaxies and found a small ensemble rotation in the same direction as the primary disks.  Their Figure 8 shows that the fraction of satellites in their sample with projected prograde orbits is $f_{\rm pro} \simeq 0.54$.  Our Equation \ref{eq:error_on_f} suggests that the counting error implies $f_{\rm pro} = 0.54 \pm 0.05$ for this measurement, which is consistent within $0.2 \sigma$ of our all-inclinations sample prediction of $0.53$. \citet{Azzaro2006} analyzed the orbital rotation directions of 76 satellite galaxies around 43 disk galaxy primaries extracted from the Sloan Digital Sky Survey (SDSS) and reported $f_{\rm pro} = 0.61 \pm 0.05$, which is $1.6 \sigma$ above our prediction. 
\citet{HerbertFort2008} combined new observations of 78 satellites around 63 isolated SDSS-DR4 host galaxies with 91 satellites around 55 hosts from \citet{Zaritsky1997}. Their combined sample contains 169 satellites around 118 host galaxies and gives a result in perfect agreement with our fiducial prediction, $f_{\rm pro}=0.53\pm0.04$. \citet{HwangPark2010} used a catalog of 635 satellites around 215 host galaxies with spin directions determined by inspection of the SDSS-DR7 color images and/or spectroscopic observations in the literature and found $f_{\rm pro}=0.50\pm0.02$.  This is $\sim 1.5 \sigma$ below our prediction. Treating the four measurements discussed above as independent estimates would suggest $f_{\rm pro}=0.520\pm0.016$, which is consistent with our prediction at $0.6 \sigma$. If Zaritsky et al. is omitted to avoid its overlap with Herbert-Fort et al., the naïve combination becomes $f_{\rm pro}=0.518\pm0.017$, which is $0.7 \sigma$ below our prediction. 

More recently, \citet{Wang2024} combined GAMA+SAMI with SDSS+MaNGA to identify $H=580$ hosts with disk rotation fields and with most systems having one or two spectroscopic satellites. 
Though they did not report spins or prograde fractions, they did report a positive co-rotation signal at $3 \sigma$. Assuming $HN = 870$, the same data set should enable an independent measurement of $f_{\rm pro}$ to within $\pm 0.017$.~\footnote{Interpreted with this uncertainty, their $3 \sigma$ claim of positive co-rotation would be consistent with $f_{\rm pro} \simeq 0.55 \pm 0.017$, which is within $\sim 1.2 \sigma$ of our fiducial prediction.}
The current SAGA DR3 sample \citep{Mao2021,Mao2024} could provide similar power, relying on fewer hosts ($H = 101$) but a deeper satellite census ($N \simeq 3.74$, to $M_\star \simeq 10^{7 - 7.5}$ M$_\odot$). Assuming disk spin orientations can be obtained for all of them, then SAGA's $HN = 378$ would enable  
 an estimate of $f_{\rm pro}$ to within $\pm 0.025$. 

Note that the fiducial $f_{\rm pro}$ prediction we have been discussing above refers explicitly to the expected average over all disk inclination angles: $0.530$. When we restrict to more highly inclined systems, $i > 45^\circ$, the prediction is slightly higher, $0.534$ (see Appendix \ref{sec:inclination} for details). In practice, the detailed dependence on inclination angle should be accounted for when we compare to observations with a relevant level of statistical precision for this third decimal point to matter.

Similar data sets would also allow for a measurement of $\lambda_\parallel$. The idealized error on $\lambda_\parallel$ quoted in Equation \ref{eq:error_on_mean} assumes perfect knowledge of the host $M_{\rm vir}$. A more realistic estimate would allow for uncertainty in virial mass based on an empirical galaxy--halo model such as UniverseMachine \citep{Behroozi2019UniverseMachine}, which would provide an estimate of $P(M_{\rm vir}|O)$, where $O$ is some set of observables including stellar mass and star formation rate. We show in Appendix \ref{app:halo_mass_uncertainty} that a 0.3 dex uncertainty at a fixed observable set would increase the error on the mean in Equation \ref{eq:error_on_mean} by a factor of about $1.1$ (see Equation \ref{eq:error_on_mean2}).  With this factor included, the SAGA DR3 sample would enable a measurement of $\lambda_{\parallel}$ to within $\pm 0.02$; a sample like that of \citet{Wang2024} would allow a measurement of $\lambda_{\parallel}$ to within $\pm 0.015$.  An extension of this approach could rely on MaNGA DR17 \citep{Abdurrouf2022} plus DESI LOW-Z \citep{DarraghFord2023} targets to measure $H \sim 1000$ disk orientations with $N \sim 2$ spectroscopic satellites each to give $\lambda_{\parallel}$ to within $\pm 0.01$.

The observational uncertainties quoted above are for idealized surveys that reach counting-statistics floors.  They do not include a realistic treatment of interlopers, satellite velocity errors, or uncertainties related to disk position angle, inclination, or receding-side assignment.   A more realistic forecast would require forward-modeling the satellite selection, velocity errors, contamination, and disk-kinematic uncertainties.  Nevertheless, there appears to be a realistic opportunity to determine whether the orbital angular-momentum content of satellite tracers, and its alignment with host-galaxy disks, are consistent with predictions of $\Lambda$CDM. Such a measurement of $\lambda_\parallel$ would provide an indirect test of the tidal-torque and anisotropic-accretion processes that generate angular momentum during structure formation in an expanding universe.

\begin{acknowledgments}

The authors thank Ben Weiner, whose talk at the 2025 Santa Cruz Galaxy Workshop inspired this work.

We acknowledge the use of LLM tools including ChatGPT and Claude to assist with coding, statistical estimates, and for checking the mathematical derivations.  We have verified all equations and analysis scripts and take full accountability for the results and the code, which is available in the provided Zenodo repository: \url{https://doi.org/10.5281/zenodo.24541135}. The archive includes measured catalogs, a Jupyter notebook, and code to regenerate the projected measurements and mock surveys.
  
LYX acknowledges funding support from the National Science Foundation Graduate Research Fellowship Program under Grant No.~DGE-2146755 and a Stanford Graduate Fellowship in Science and Engineering. JSB is supported by NSF grant AST-2408246.

We have used simulations from the Auriga Project public data release \citep{Grand2017,Grand2024}. We also use the publicly available FIRE-2 cosmological zoom-in simulations \citep{Wetzel2023,Wetzel_2025_DR2} from the Feedback In Realistic Environments (FIRE) project, generated using the \textsc{gizmo} code \citep{Hopkins2015} and the FIRE-2 physics model \citep{Hopkins2018}. We also use publicly available data from IllustrisTNG
\citep{Nelson_2019}, including the TNG50 simulation
\citep{Pillepich2019TNG50,Nelson2019TNG50}, and from the Symphony simulation suites \citep{nadler2023}.
\software{NumPy, SciPy, pandas, Matplotlib, seaborn, h5py, symlib}
\end{acknowledgments}

\clearpage
\appendix

\section{Measuring Projected Satellite Angular Momentum}
\label{app:projected-covariance}

In Section \ref{sec:observational} we introduce an observable that allows us to estimate the projected satellite spin:
\begin{equation}
    \ell_s \equiv  \frac{2x \Delta v_{\rm los}}{\sqrt{2}R_{\rm vir}V_{\rm vir}\sin i}.
    \label{eq:ell_general_orientation}   
\end{equation}
Here we demonstrate that the expectation value for this estimator is $\langle \lambda_\parallel \rangle$ (where $\lambda_\parallel$ is  defined in Equation \ref{eq:lambda_par}) and further that the variance in $\ell_s$ within any given halo $h$ is $\sigma_{\ell, h}^2 = C_{\rm proj} \lambda_{{\rm sat}}^{2}\kappa_{{\rm sat}}$, where $C_{\rm proj} = 2/ (3\sin^2i)$ and $\kappa_{\rm sat}$ is the satellite angular momentum contrast for halo $h$, as defined in Equation \ref{eq:kappa-sat}.

Figure \ref{fig:viewing_example} shows our observational setup. We let the disk angular-momentum vector point along the $z$-axis, and let the $x$-axis coincide with the projected major axis of the disk.  The line of sight lies in the $y$--$z$ plane. With this geometry, the line-of-sight velocity $\Delta v_{\rm los}$ connects to the 3D velocities of satellites as 
\begin{equation}
    \Delta v_{\rm los}= v_y\sin i+ v_z\cos i,
\end{equation}
where $i$ is the inclination of the disk on the sky such that $i=90^\circ$ is edge on.  Our primary observable is $x \Delta v_{\rm los}$, which has units of specific angular momentum.

Now define
\begin{equation}
    a\equiv xv_y, \quad  b\equiv yv_x, \quad c\equiv xv_z.
\end{equation}
With these definitions, 
our observable is
\begin{align}
 x \Delta v_{\rm los} = a \sin i +c  \cos i.
\label{eq:ell_general_orientation2}
\end{align}
Compare this to the true satellite angular momentum component along the disk axis:
\begin{equation}
j_z=a-b.
\end{equation}  
For a statistically symmetric satellite population  $\langle a \rangle = -\langle b \rangle$, which implies $\langle j_z \rangle = 2 \langle a \rangle$.  Our geometric setup further demands $\langle c\rangle=0$, which implies
\begin{equation}
\langle x \Delta v_{\rm los} \rangle
=  \frac{ \langle j_z\rangle \sin i}{2}.
\label{eq:observable}
\end{equation}
For fixed virial parameters, this is equivalent to our first key result:
\begin{equation}
\boxed{
    \langle \ell_s \rangle = \langle \lambda_{\parallel}\rangle.}
\end{equation}
Here, $\lambda_{\parallel} = \lambda_z$ is the dimensionless satellite spin projected onto the direction of the disk spin axis, $z$.

Even though $\ell_s$ is an unbiased estimator of $\lambda_{z}$, the variance in the observable $\ell_s$ within any given halo is larger than the variance in $\lambda_z$. To see this, assume that the relevant second moments are isotropic:
\begin{equation}
    \sigma_a^2  = \sigma_b^2  = \sigma_c^2  \equiv s^2.
\end{equation}
Further assume no covariance between $a, b,$ and $c$. This allows us to write the variance in our observable $x \Delta v_{\rm los}$ as
\begin{equation}
    \sigma^2_{x \Delta v_{\rm los}} = \sigma_a^2  \sin^2 i + \sigma_c^2  \cos^2i = s^2 ,
    \label{eq:obs_variance_general}
\end{equation}
while the variance in $j_z$ is
\begin{equation}
    \sigma^2_{j_z} = \sigma_a^2 + \sigma_b^2 = 2s^2.
\end{equation}
The associated variance in $\lambda_z$ among satellites of a given halo $h$ is
\begin{align}
    \sigma^2_{\lambda_z,h} = \frac{2 s^2}{2 R^2_{\rm vir} V^2_{\rm vir}},
    \label{eq:lambda_variance_general}
\end{align}
whereas the variance in $\ell_s$ is larger
\begin{equation}
    \sigma^2_{\ell_s,h} = \frac{4 s^2}{2 R^2_{\rm vir} V^2_{\rm vir} \sin^2 i} = \frac{2 }{\sin^2 i} \sigma^2_{\lambda_z,h}.
\label{eq:ell_variance_lambdaz}
\end{equation}

Under isotropy, the variance in the full three-dimensional angular-momentum vector among satellites of a halo is equally distributed among its three components:
\begin{equation}
    \sigma^2_{\boldsymbol{\lambda}3D} = 3 \sigma^2_{\lambda_z} = \lambda_{\rm sat}^2\kappa_{\rm sat},
\end{equation}
where the last equality uses Equation \ref{eq:lambda_var}. Combining this with Equation \ref{eq:ell_variance_lambdaz} yields our second key result
\begin{equation}
    \boxed{
    \sigma_{\ell,h}^2 = \frac{2}{3\sin^2i}\, \lambda_{{\rm sat}}^2\kappa_{{\rm sat}}.
    }
    \label{eq:sigma_ell_final}
\end{equation}
Averaging over angles, restricting to those with $i > 45^\circ$, gives an average prefactor 
\begin{equation}
    C_{\rm proj} =  \frac{2}{3} \left\langle\frac{1}{\sin^2 i}\right\rangle \simeq 0.83.
\end{equation}

In the main text, we explore scenarios where we draw multiple satellite populations viewed from random angles to test Equation \ref{eq:sigma_ell_final} directly and show that it works quite well. 
In a subset of the mock experiments we perform in the text, we utilize a fixed set of satellites, which we view from a large number of orientations. The orientation averaging recovers its true \(\lambda_z\), which removes the \(\sigma_{\lambda_z}^{2}\) contribution from the full variance $\sigma^2_\ell$. The expected   ``orientation-only" contribution to the variance is
\begin{equation}
\sigma_{\rm orient,h}^{2} = \sigma_{\ell,h}^{2}-\sigma_{\lambda_z,h}^{2} = \left(1-\frac{\sin^2 i}{2}\right) \sigma_{\ell,h}^{2},
\label{eq:orientation_variance}
\end{equation}
which is a bit smaller than the full variance we would want to include for survey planning. For example, assuming an isotropic distribution of viewing angles restricted to those with $i > 45^\circ$ we expect $\sigma_{\rm orient} \simeq 0.774 \sigma_\ell$, which is roughly what we observe in the relevant mocks.

\section{Co-rotating fractions}
\label{App:co-rot}

Adopting the geometry of Figure \ref{fig:viewing_example}, 
a satellite is classified as prograde when its projected position and line-of-sight velocity have the same sign, such that the projected prograde fraction is

\begin{equation}
    f_{\rm pro} \equiv P\!\left(x_s\Delta v_{{\rm los},s}>0\right).
    \label{eq:prograde-definition}
\end{equation}

Assuming an observational program involving $H$ hosts with $N$ satellites each, the accuracy with which we can measure the population-level $\langle f_{\rm pro} \rangle$ will depend on the
variance in the prograde fraction among satellites within a given host, $\sigma_{f,h}^2$, and the host-to-host variance $\sigma_{\rm f_{\rm pro}}$:
\begin{equation}
    \sigma^2_{\langle f_{\rm pro}\rangle} = \frac{1}{H}\left[ \sigma_{f_{\rm pro}}^2 + \frac{\langle \sigma^2_{f,h}\rangle}{N} \right].
\end{equation}
Since prograde/retrograde is a binary variable, the average variance in the prograde fraction within a halo is $\langle \sigma_{f,h}^2\rangle = {\left\langle f_{{\rm pro},h} (1-f_{{\rm pro},h})\right\rangle_h} \simeq 1/4$.
In our simulations, we measure $\sigma_{f_{\rm pro}}^2 \sim 10^{-3}$, which implies that the second term dominates for $N \lesssim 100$:
\begin{equation}
    \boxed{
    \sigma_{\langle f_{\rm pro}\rangle} \simeq \frac{1}{2\sqrt{HN}}.}
\end{equation}

We expect the projected prograde fraction to be correlated with the projected spin. In order to see this, assume that $x_s$ and $\Delta v_{{\rm los},s}$ are centered, so that their correlation coefficient is

\begin{equation}
\rho = \frac{\left\langle x\Delta v_{\rm los}\right\rangle}
{\sigma_x\sigma_v}.
\label{eq:rho-projection-appendix}
\end{equation}

We can then make the simplifying assumption that both $x_s$ and $\Delta v_{los,s}$ are jointly Gaussian variables that have within-host dispersions $\sigma_x$ and $\sigma_v$.  In this case, the probability that both variables have the same sign is given by Sheppard’s bivariate quadrant probability formula \citep{Tong1990} 

\begin{equation}
    f_{\rm pro}=\frac{1}{2}+\frac{1}{\pi}\arcsin(\rho).
    \label{eq:prograde-gaussian}
\end{equation}

For $|\rho|\ll1$, we may approximate $\arcsin(\rho)\simeq\rho$ and use Equation \ref{eq:observable} to write
\begin{equation}
    f_{\rm pro}\simeq\frac{1}{2}+A\lambda_\parallel\sin i. 
    \label{eq:fpro-approx-appendix}
\end{equation}
Here, the coefficient $A = R_{\rm vir}V_{\rm vir}/{\pi\sqrt{2}\,\sigma_x\sigma_v}$. This quantity can vary from host to host but is expected to be of order unity. 

\begin{figure}[t!]
    \centering
    \includegraphics[width=\linewidth]
    {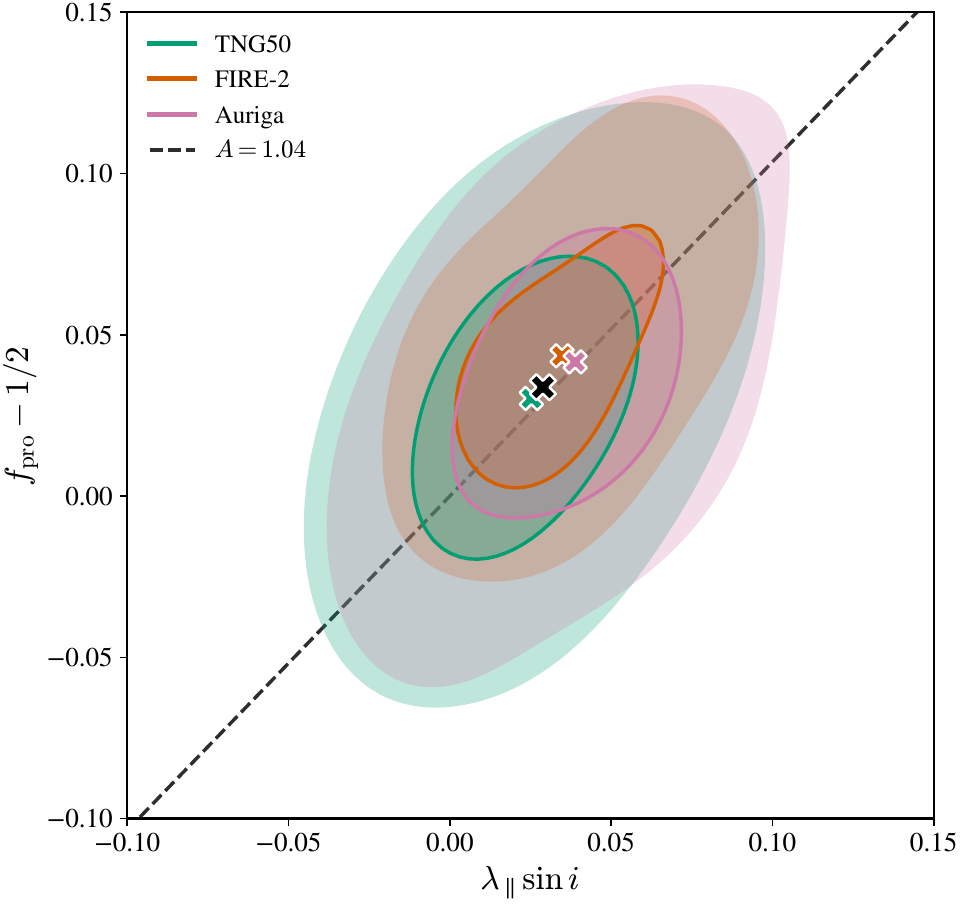}
    \caption{Test of Equation \ref{eq:fpro-approx-appendix}, plotting measured prograde excess, $f_{\rm pro}-1/2$, versus $\lambda_\parallel\sin i$, where $\lambda_\parallel=\lambda_{\rm sat}\cos\theta_{\rm sat,disk}$. Light and dark smoothed-density regions are set to contain 90\% and 50\%, respectively, of each suite's host--sightline measurements. The x's mark the suite means and the black x marks the combined mean. The dashed line is the empirical zero-intercept fit $f_{\rm pro}-1/2=1.04\lambda_\parallel\sin i$, with a 95\% host-bootstrap interval of $0.95$--$1.11$ on the slope.}
    \label{fig:fpro-validation}
\end{figure}

Figure \ref{fig:fpro-validation} tests the expectation in Equation \ref{eq:fpro-approx-appendix}.  We measure $f_{\rm pro}$ by viewing each host in our fiducial hydrodynamic sample 400 times from random orientations. We only keep orientations with $i\geq45^\circ$. The resultant probability densities are color coded by each simulation type, the light and dark filled contours enclose 90\% and 50\%, respectively, of the host--sightline measurements, and the x's mark the averages of $f_{\rm pro} - 1/2$ and  $\lambda_\parallel \sin i$ for each sample.

The prograde fraction increases with $\lambda_\parallel \sin i$ in a roughly linear fashion.  The dashed line is Equation \ref{eq:fpro-approx-appendix} with the value of $A$ fitted to $1.04$. 

\section{Effect of uncertainties in $M_{\rm vir}$ }
\label{app:halo_mass_uncertainty}

Determining $\lambda_\parallel$ using Equation~ \ref{eq:projected-satellite-contribution} requires estimates of $R_{\rm vir}$ and $V_{\rm vir}$, where $\lambda_\parallel \propto 1/R_{\rm vir}V_{\rm vir}\propto M_{\rm vir}^{-2/3}$. The relevant uncertainty is therefore related to the distribution $P(M_{\rm vir}\mid O)$, where $O$ is some set of observables. For example, empirical galaxy--halo models such as UniverseMachine \citep{Behroozi2019UniverseMachine} can be used to construct this distribution by selecting mock galaxies with the same observed properties, such as stellar mass, star-formation rate, and central-galaxy status.  The observed rotation velocity would also likely provide additional information. Here, we provide an illustrative estimate of the additional uncertainty that imperfect knowledge of the halo normalization introduces into measurements of $\langle\lambda_\parallel\rangle$.  A complete analysis would forward-model the observational selection and marginalize over $P(M_{\rm vir}\mid O)$ throughout the measurement.

For each host, let $R_{{\rm vir},h}$ and $V_{{\rm vir},h}$ represent its true virial radius and virial velocity, both of which are not observed.  For this analysis, we assume that an observable $O$ is used to infer these quantities. Let $\langle R_{\rm vir}V_{\rm vir}\rangle_{O}$ be the mean inferred spin normalization implied by the adopted $P(M_{\rm vir}\mid O)$ distribution. With this, we can define a multiplicative normalization error
\begin{equation}
w_h\equiv
\frac{R_{{\rm vir},h}V_{{\rm vir},h}}{\langle R_{\rm vir}V_{\rm vir}\rangle_{O,h}},
\end{equation}
such that the expectation value for the projected satellite spin of this halo is
\begin{equation}
    \mathbb{E}\left[\lambda_{N,h}\right] = w_h \lambda_{\parallel,h}.
\end{equation}
Although every halo has one fixed value of $w_h$, its value is unknown observationally. 
We have defined $w$ so that over many halos we expect $\langle w \rangle = 1$, which means that the full stack should still provide an unbiased estimator
\begin{equation}
    {\rm E}\left[ \widehat{\lambda}_{\rm stack}\right] = \langle \lambda_\parallel \rangle.
    \label{eq:mass_weight_expectation}
\end{equation}
However, the implied variance in the inferred spin of each halo ($w_h \lambda_{\parallel,h}$) necessarily increases.  Specifically, the second term in Equation \ref{eq:variance_stack1} is replaced by $\langle \sigma^2_{\lambda_{N,h}}\rangle \rightarrow \sigma^2_{w \lambda_\parallel}$, where
\begin{equation}
   \sigma^2_{w \lambda_\parallel} = \langle w^2\rangle \langle \sigma^2_{\lambda_{N,h}}\rangle+ \sigma_w^2 \langle \lambda_{\parallel}^2 \rangle.
   \label{eq:wlambda_var}
\end{equation}
Here, $\langle \sigma^2_{\lambda_{N,h}}\rangle = \langle \lambda_{\parallel} \rangle^2 \kappa_\parallel/N$ is the sampling variance in the projected satellite spin with $N$ satellites. For realistic $N$, the first term in Equation \ref{eq:wlambda_var} dominates.  Now when the experiment is done using $H$ hosts, the total variance of our stacked estimator (Equation \ref{eq:error_on_mean}) is similar to that in the idealized case, modulo a factor of $\langle w^2 \rangle$:
\begin{equation}
    \sigma_{\langle\lambda_\parallel\rangle} = \sqrt{\frac{ \langle w^2 \rangle \langle \lambda_{\parallel} \rangle^2 \kappa_\parallel}{HN}} \simeq 0.4 \sqrt{\frac{\langle w^2 \rangle}{HN}}.
    \label{eq:error_on_mean2}
\end{equation}
This means the total sample size, $HN$, can be increased by a factor of $\langle w^2\rangle$ to achieve the same precision. If $w$ is a mean-normalized log-normal variable, we know its second moment is related to its log scatter $\langle w^2\rangle = \exp(\sigma_{\ln w}^2)$. 

Observational constraints indicate substantial scatter in the galaxy--halo connection near Milky-Way halo masses; for example, \citet{Cao2020} infer a forward scatter $\sigma_{\log_{10}M_\star\mid M_{\rm h}}\simeq0.34$--$0.38$ dex. The inverse scatter relevant here, $\sigma_{\log_{10}M_{\rm vir}\mid O}$, will depend on the slope of the stellar-to-halo mass relation and the halo-mass prior. The sample we envision for our experiments will include only central galaxies with star-forming, rotating gas disks, and this would reduce the inverse scatter by removing the passive-galaxy population, which occupies systematically more massive halos at fixed stellar mass \citep[e.g.][]{More2011SatelliteKinematics,Mandelbaum2016Bimodality}. As an illustrative model, we assume $\sigma_{\log_{10}}=0.30$ dex, which implies   $\sigma_{\ln w} = 2/3 \, \ln(10)\sigma_M =0.46$ and $\langle w^2\rangle =1.24.$  In this case,  virial-mass scatter increases the mean error by $\sim 11\%$ and the required satellite count would need to be increased by $24\%$ to maintain the same precision.  In practice, the inverse distribution should ultimately be obtained by applying the observational host selection, including its disk or star-forming selection, to a forward galaxy--halo model.

\section{Decomposing the RMS Angular-Momentum Ratio}
\label{app:rj}
In subsection \ref{subsec:whybigger} we mentioned that larger satellite orbital radii are the primary reason for $\lambda_{\rm sat} > \lambda_{\rm dm}$. We explain why as follows.

For either population $X\in\{{\rm sat},{\rm dm}\}$, define
$r_{\rm rms,X}=\langle r^2\rangle_X^{1/2}$ and
$v_{\rm rms,X}=\langle v^2\rangle_X^{1/2}$.
The specific angular momentum of either population can be written as $j_{\rm rms,X}=r_{\rm rms,X}v_{\rm rms,X}G_X^{1/2}$, where $G$ is an orbital geometric factor
\begin{equation}
    G_X\equiv\frac{\langle|\boldsymbol r\times\boldsymbol v|^2\rangle_X}{\langle r^2\rangle_X\langle v^2\rangle_X}.
\end{equation}
Larger values of $G$ reflect larger fractions of tangential motion. With this definition, the ratio in Equation~\ref{eq:spin_ratio_split} separates exactly as
\begin{equation}
    R_j={} \frac{r_{\rm rms,sat}}{r_{\rm rms,dm}} \frac{v_{\rm rms,sat}}{v_{\rm rms,dm}} \left(\frac{G_{\rm sat}}{G_{\rm dm}}\right)^{1/2}.
    \label{eq:moment-factorization}
\end{equation}
For the 176 fiducial hosts, we measure the median radius ratio, the median velocity ratio, and the median geometry ratio to be
\begin{equation}
\begin{aligned}
\frac{r_{\rm rms,sat}}{r_{\rm rms,dm}}
    &= 1.39^{+0.12}_{-0.17},\\
\frac{v_{\rm rms,sat}}{v_{\rm rms,dm}}
    &= 1.00^{+0.06}_{-0.04},\\
\left(\frac{G_{\rm sat}}{G_{\rm dm}}\right)^{1/2}
    &= 1.13^{+0.09}_{-0.13}.
\end{aligned}
\end{equation}
The ranges span the 16th--84th
percentiles across hosts. Thus, we conclude that larger satellite orbital radii provide the main contribution to $R_j$; the geometric factor adds a smaller enhancement; the velocity bias is largely unimportant. 

\section{Effects of Inclination on Prograde Fraction}
\label{sec:inclination}

\begin{figure}[t!]
    \centering
    \includegraphics[width=\linewidth]{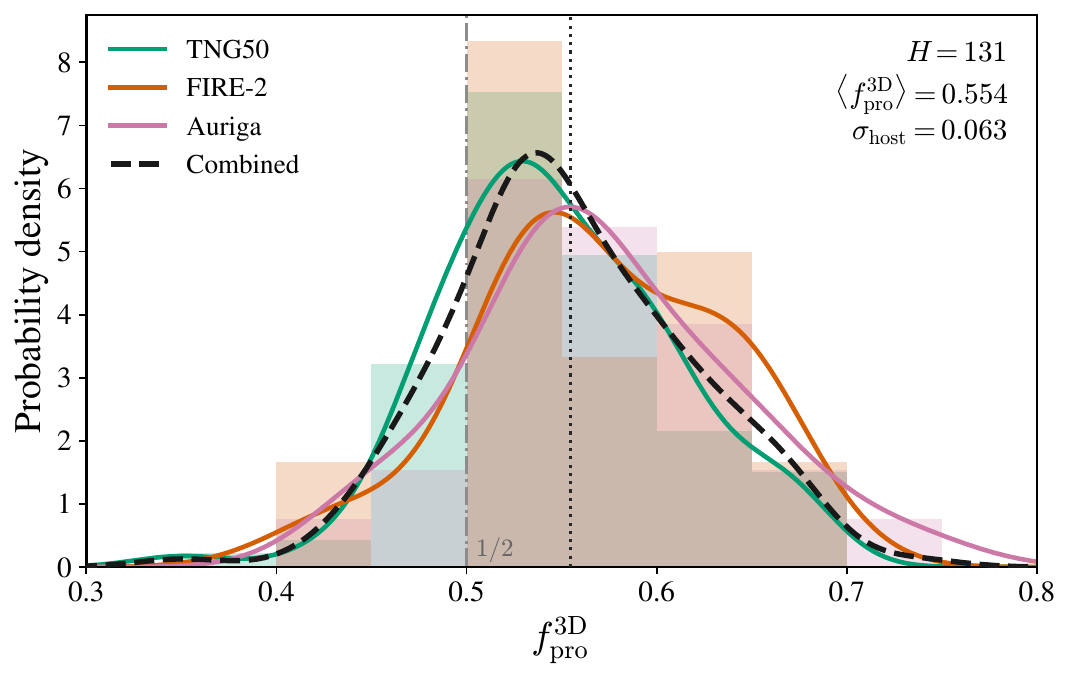}
    \caption{Distribution of 3D prograde fractions, $f_{\rm pro}^{3D}$, for our fiducial sample of hydrodynamic hosts.  This is the fraction of a given host's satellites that have $(\mathbf r_s\times\mathbf v_s) \cdot\widehat{\mathbf j}_{\rm disk}>0$, which requires full 3D information.}
    \label{fig:fpro-3d}
\end{figure}

The intrinsic three-dimensional prograde fraction is defined by the sign of the satellite orbital angular momentum relative to the host-disk spin: P($(\mathbf r_s\times\mathbf v_s)\cdot\widehat{\mathbf j}_{\rm disk}>0$). As shown in Figure \ref{fig:fpro-3d}, we find $\langle f_{\rm pro}^{\rm 3D}\rangle =0.554$, when averaged over all hosts.

The projected observable, $f_{\rm pro}$, instead measures $P(x_s\Delta v_{{\rm los},s} > 0)$. Because $x_s\Delta v_{{\rm los},s}$ contains only projected position and velocity information there will be some degradation of the 3D signal.  If $\epsilon_+(i)$ and $\epsilon_-(i)$ denote the probabilities of misclassifying intrinsically prograde and retrograde satellites, respectively, then we can write
\begin{equation}
    f_{\rm pro}(i) = f_{\rm pro}^{\rm 3D}\,[1-\epsilon_+(i)] + [1-f_{\rm pro}^{\rm 3D}]\,\epsilon_-(i).
\end{equation}
For approximately symmetric classification errors, $\epsilon_+\simeq\epsilon_-\equiv\epsilon$, this becomes
\begin{equation}
    f_{\rm pro}(i) \simeq \frac{1}{2} + [1-2\epsilon(i)] \left(f_{\rm pro}^{3D}-\frac{1}{2}\right).
\end{equation}

For $\sin i = 0$, we expect to lose all signal and find $f_{\rm pro} = 1/2$.  For $\sin i = 1$, the disk will be perfectly edge on and we will be able to infer prograde rotation with the least amount of error. This suggests that $[1-2\epsilon(i)] \propto \sin i$, and motivates us to guess
\begin{equation}
    f_{\rm pro}(i) \simeq \frac{1}{2} + \eta \sin i \left(f_{\rm pro}^{\rm 3D}-\frac{1}{2}\right),
\end{equation}
where $\eta$ is an empirical factor. Using $f_{\rm pro}^{\rm 3D} = 0.554$, this implies
\begin{equation}
    f_{\rm pro}(i) \simeq \frac{1}{2} + 0.054 \, \eta \sin i.
\end{equation}

\begin{figure}[t!]
    \centering
    \includegraphics[width=\linewidth]{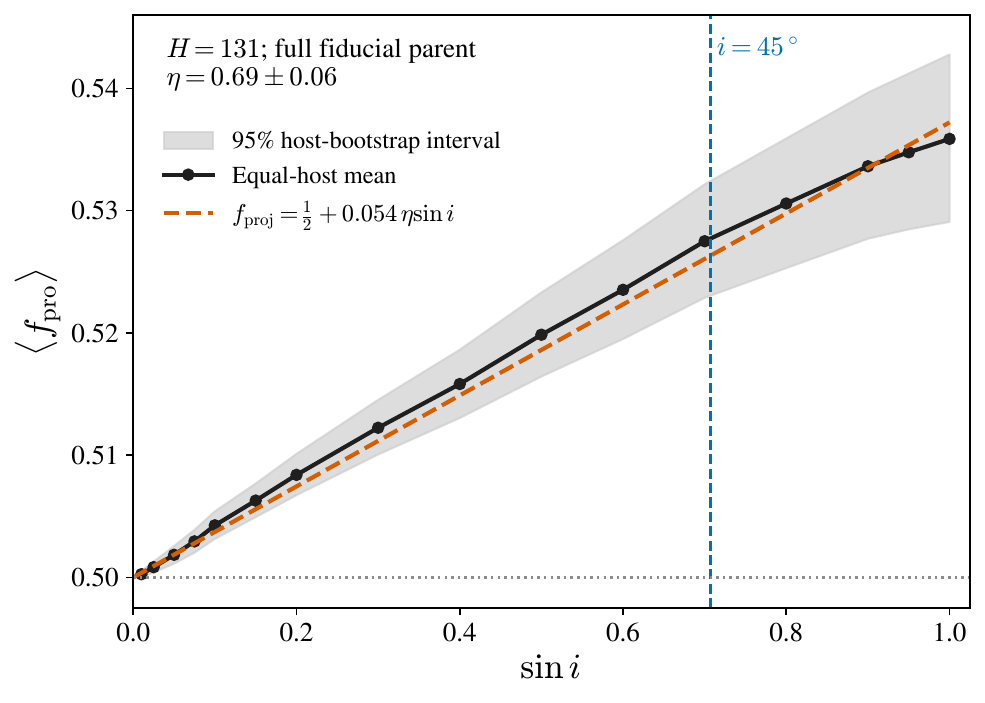}
    \caption{Dependence of the measured projected prograde fraction as a function of inclination angle of the disk using our fiducial hydrodynamic sample. The dashed line shows Equation \ref{eq:fpro_sini}.}
    \label{fig:fpro-sini}
\end{figure}

Using mock surveys of our simulated halos, we find $\eta \simeq 0.69$ and this allows us to write
\begin{equation}
    f_{\rm pro}(i) \simeq 0.5+0.037\sin i .
    \label{eq:fpro_sini}
\end{equation}
This line is plotted in Figure \ref{fig:fpro-sini}, where we show the observed prograde fraction as a function of inclination angle of the simulated disk. An isotropic sample has $\langle\sin i\rangle=\pi/4$, implying $f_{\rm pro} = 0.530$; restricting to high inclinations gives $\langle\sin i\rangle_{i\ge45^\circ} = 0.909$, implying  $f_{\rm pro} = 0.534$.

Note that the linear projected-spin estimator, $\lambda_\parallel$ does not depend on the distribution of inclination angles in the data because inclination produces a deterministic factor, \(\langle\lambda_{\rm proj}\mid i\rangle= 0.5 \lambda_\parallel\sin i\), which is removed by multiplying by \(2/\sin i\). Instead, the inclination  matters for the {\em variance} in \(\lambda_\parallel\) estimates through the \(1/\sin^2 i\) factor in $C_{\rm proj}$ defined within Equation \ref{eq:error_on_mean}.

\newpage
\bibliographystyle{aasjournalv7}
\bibliography{ref}
\end{document}